\documentclass[review]{elsarticle}

\usepackage{lineno,hyperref}
\usepackage{multirow}
\usepackage{tabularx}
\usepackage{caption}
\usepackage{subfigure}
\usepackage{diagbox}
\usepackage{pgfplots}
\usepackage{longtable}
\usepackage{pgfplotstable}
\usepackage{pdflscape}

\usepackage{cleveref}
\usepackage{makecell}

\begin{document}

\begin{frontmatter}

\title{Automatic Diagnosis of COVID-19 from CT Images using CycleGAN and Transfer Learning}

\author[labkntu,ferdowsi]{Navid Ghassemi}
\cortext[data]{All data and codes for this work have been made available publicly on the following link: \href{https://github.com/afshin0919/Shoeibi-COVID19-Dataset}{github.com/afshin0919/Shoeibi-COVID19-Dataset}}
\author[labkntu,ferdowsi]{Afshin Shoeibi\corref{mycorrespondingauthor}}
\cortext[mycorrespondingauthor]{Corresponding author}
\ead{afshin.shoeibi@gmail.com}
\author[mhd]{Marjane Khodatars}

\author[har]{Jonathan Heras}
\author[ferdowsi]{Alireza Rahimi}
\author[ass]{Assef Zare}
\author[pac]{Ram Bilas Pachori}
\author[rem]{J. Manuel Gorriz}

\address[labkntu]{Faculty of Electrical Engineering, Clinical Studies Lab, K. N. Toosi University of Technology, Tehran, Iran.}
\address[ferdowsi]{Computer Engineering department, Ferdowsi University of Mashhad, Mashhad, Iran.}
\address[mhd]{Department of Medical Engineering, Mashhad Branch, Islamic Azad University, Mashhad, Iran.}
\address[har]{Department of Mathematics and Computer Science, University of La Rioja, La Rioja, Spain.}
\address[ass]{Faculty of Electrical Engineering, Gonabad Branch, Islamic Azad University, Gonabad, Iran.}
\address[pac]{Department of Electrical Engineering, Indian Institute of Technology Indore, Indore 453552, India.}
\address[rem]{Department of Signal Theory, Networking and Communications, Universidad de Granada, Spain.}

\begin{abstract}
The outbreak of the corona virus disease (COVID-19) has changed the lives of most people on Earth. Given the high prevalence of this disease, its correct diagnosis in order to quarantine patients is of the utmost importance in steps of fighting this pandemic. Among the various modalities used for diagnosis, medical imaging, especially computed tomography (CT) imaging, has been the focus of many previous studies due to its accuracy and availability. In addition, automation of diagnostic methods can be of great help to physicians. In this paper, a method based on pre-trained deep neural networks is presented, which, by taking advantage of a cyclic generative adversarial net (CycleGAN) model for data augmentation, has reached state-of-the-art performance for the task at hand, i.e., 99.60\% accuracy. Also, in order to evaluate the method, a dataset containing 3163 images from 189 patients has been collected and labeled by physicians. Unlike prior datasets, normal data have been collected from people suspected of having COVID-19 disease and not from data from other diseases, and this database is made available publicly.
\end{abstract}


\end{frontmatter}

\section{Introduction}
The COVID-19 disease was initially spotted in December of 2019 in Wuhan, China, and was detected worldwide shortly after that \cite{i1}. In January 2020, the World Health Organization (WHO) stated its outbreak as a public health emergency and global concern, and later, a pandemic in March of 2020 \cite{i2}. SARS-CoV-2 causes COVID-19, a novel variety of coronavirus that has not been identified beforehand in humans \cite{i3i4}. Coronaviruses are common among animals, and some can infect humans \cite{i3i4,i5}. Bats are the natural hosts of these viruses, and several other species of animals have also been identified as sources \cite{i6,i7,nnone}. For example, MERS-CoV3 is transmitted from camels to humans, while SARS-CoV-14 is transmitted from intermediate hosts such as civet cats that were involved in the development of SARS-CoV-1 \cite{i8,i9}. The new coronavirus is genetically closely related to the SARS-CoV-1 virus \cite{i10}.

The SARS-CoV2 virus is transmitted mainly through respiratory droplets and aerosols from an infected person while sneezing, coughing, talking, or breathing in the presence of others \cite{i11,i12}. The virus can survive at varying surfaces from a few hours to several days, and prior research has estimated the incubation period of this disease to be within 1 and 14 days \cite{i13}. However, the amount of live virus decreases over time and may not always be present in sufficient quantities to cause infection \cite{i14,i15}.

The most frequent symptoms of COVID-19 include fever, dry cough, and fatigue. Pain, diarrhea, headache, sore throat, conjunctivitis, loss of taste or smell are other variable symptoms of the virus \cite{i16,i17}. Nevertheless, the most severe symptoms seen in COVID-19 patients include difficulty for breathing or shortness of breath, chest pain, and loss of movement or ability to speak \cite{i16,i17,i18}.

Early diagnosis of this disease in the preliminary stages is vital. So far, various screening methods have been introduced for the diagnosis of COVID-19. At present, nucleic acid-based molecular diagnosis (RT-PCR5 test) is considered the gold standard for early detection of COVID-19 \cite{i19,i20}. According to a WHO report, all diagnoses of COVID-19 must be verified by RT-PCR \cite{i21}. However, performing the RT-PCR test needs specialized equipment and equipped laboratories that are not available in most countries and takes at least 24 hours to determine the test outcome. Also, the test result may not be accurate and may require re-RT-PCR or other tests. Therefore, X-ray and CT-Scan imaging can be used as a primary diagnostic method for screening people suspected of having COVID-19 \cite{i22,i23}.

X-ray imaging is one of the medical imaging techniques used to diagnose COVID-19. X-ray imaging benefits include low cost and low risks of radiation that are dangerous to human health \cite{i24,i25}. In this imaging technique, the detection of COVID-19 is a relatively complicated task. An X-ray physician may also misdiagnose diseases such as pulmonary tuberculosis \cite{i26,i27}.

CT-Scan imaging is used to reduce COVID-19 detection error. CT-scans have very high contrast and resolution, and are very successful in diagnosing lung diseases such as COVID-19 \cite{i28,i29}. CT-Scan can also be used as a clinical feature of COVID-19 disease patients. CT scans of subjects with COVID-19 had shown marked destruction of the pulmonary parenchyma 6, such as interstitial inflammation and extensive consolidation \cite{i30}. During CT-Scan imaging of patients, multiple slices are recorded to diagnose COVID-19. This high number of CT-Scan images requires a high accuracy from specialists for accurate diagnosis of COVID-19. Factors such as eye exhaustion or a massive number of patients to interpret CT-Scan may lead to misdiagnosis of COVID-19 by specialists \cite{i31}.

Due to the stated challenges, the use of artificial intelligence (AI) methods for accurate diagnosis of COVID-19 on CT-Scan or X-Ray imaging modalities is of utmost importance. The design of computer aided diagnosis systems (CADS) based on AI using CT-Scan or X-Ray images for precise diagnosis of COVID-19 has been highly regarded by researchers \cite{i32,i33,abc4}. Deep learning (DL) is one of the fields of AI, and many research papers have been published on their application for diagnosing COVID-19 \cite{i34,i35}.

\begin{figure}[h]	
\includegraphics[width=\textwidth]{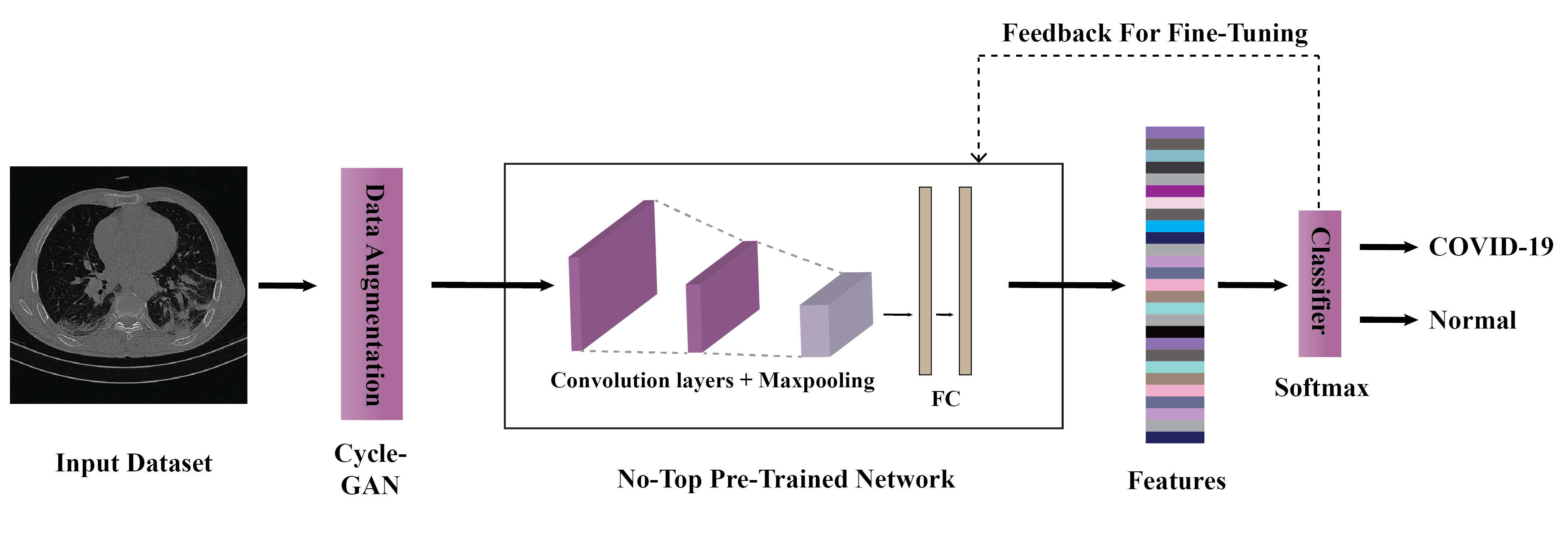}
\caption{Overall diagram of proposed method.\label{figone}}
\end{figure}

In this paper, a new method of diagnosing COVID-19 from CT-Scan images using DL is presented. First, CT-Scan images of people with COVID-19 and normal people were recorded in Gonabad Hospital (Iran). Next, three expert radiologists have labeled the patients' images. They have also selected informative slices from each scan. Then, after preprocessing data with a Gaussian filter, various deep learning networks were trained in order to separate COVID-19 from healthy patients. In this step, a CycleGAN \cite{r10,abc5} architecture was first used for data augmentation of CT-Scan data; after that, a number of pre-trained deep networks \cite{abc3} such as DenseNet \cite{r4}, ResNet \cite{r2}, ResNest \cite{r5}, and ViT \cite{r6} have been used to classify CT-Scan images. Figure \ref{figone} shows the block diagram of method. The results show that the proposed method of this study has promising results in detecting COVID-19 from CT-scan images of the lung.

The rest of the paper is organized as follows. In the next section, we present a review of previous research on the diagnosis of COVID-19 from CT-Scan images using DL techniques. In Section 3, the proposed method of this research is presented. In Section 4, the evaluation process and the results of the proposed method are presented. Section 5 includes the discussion of paper and finally, the paper ends with the conclusion and future directions. 

\section{Related Works}
 
\begin{figure}[h]	
\includegraphics[width=\textwidth]{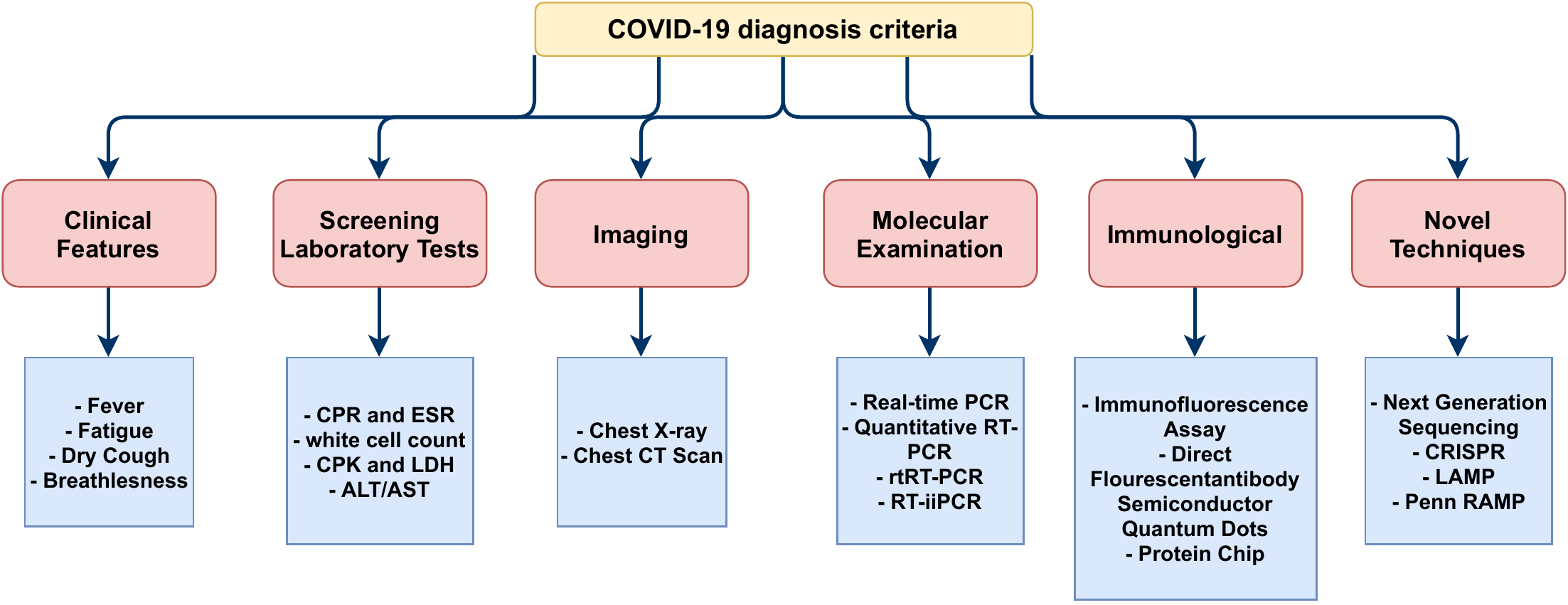}
\caption{Various criteria used for COVID-19 detection and their categories.\label{figab}}
\end{figure}

Prior research papers on the diagnosis of the COVID-19 disease using machine learning can be divided according to the algorithms used or the underlying modalities. Figure \ref{figab} shows various types of methods that can be used for diagnosis of COVID-19. As can be seen in this figure, the methods based on medical imaging can be divided into two groups: CT scan and X-ray. The focus of this article is on CT scan modality. Also, machine learning algorithms can be divided into two categories: DL \cite{ndeep,abc2} and conventional machine learning methods \cite{nbishop,nntwo,abc1}. Due to the large number of machine learning papers for diagnosing COVID-19 disease from CT modality, we have only reviewed papers that have used deep learning methods for this imaging modality. Table \ref{tab1} provides an overview of these papers, the datasets used by them, the components of methods, and finally, their performance.

\clearpage

\begin{center}
\tiny
\setlength\LTleft{-123pt}            
\setlength\LTright{0pt}           
\begin{longtable}{|c|c|c|c|c|c|c|c|c|c|}

\caption{\centering Review of related works.}
    \label{tab1}\\
    \hline
    \textbf{Ref}	& \textbf{Dataset}	& \textbf{Modality}	 & \textbf{Number of Cases}	& \textbf{Pre-Processing}	& \textbf{DNN} & \textbf{Post-Processing}	& \textbf{Toolbox}	& \textbf{K Fold}	& \makecell{ \textbf{Performance}\\\textbf{Criteria}}\\
\hline
\cite{a1}	&Clinical 	&CT	&\makecell{3000 COVID-19 Images,\\3000 Non-COVID-19 Images}	&\makecell{Patches\\Extraction}	&\makecell{VGG-16,\\GoogleNet,\\ResNet-50}	&\makecell{Feature Fusion,\\Ranking\\Technique,\\SVM}	&--	&--	&\makecell{Acc=98.27\\Sen=98.93\\Spec=97.60\\Prec=97.63}\\
\hline
\cite{a2}	&\makecell{Datasets from\\ \cite{a17} \& \cite{a8}} 	&CT	&\makecell{460 COVID-19 Images,\\397 Healthy Control (HC) Images}	&\makecell{Data\\Augmentation\\(DA)}	& \makecell{CNN\\Based on \\SqueezeNet}	&\makecell{Class Activation\\Mapping (CAM)}	&Matlab 2020a	&10	&\makecell{Acc=85.03\\Sen=87.55\\Spec=81.95\\Prec=85.01}\\
\hline
\cite{a3}	&Various Datasets	&CT	&\makecell{2373 COVID-19 Images,\\2890 Pneumonia Images,\\3193 Tuberculosis Images,\\3038 Healthy Images}	&--	&\makecell{Ensemble\\DCCNs}	&--	&Matlab 2020b	&--	&\makecell{Acc=98.83\\Sen=98.83\\Spec=98.82\\F1-Score=98.30}\\
\hline
\cite{a4}	&Clinical	&CT	&\makecell{98 COVID-19 Patients,\\103 Non-COVID-19 Patients}	&\makecell{Visual\\Inspection}	&BigBiGAN	&--	&TensorFlow	&--	&\makecell{Sen=80\\Spec=75}\\
\hline
\cite{a5}	&Clinical	&CT	&\makecell{148 Images from 66 COVID-19\\Patients, 148 Images from\\66 HC Subjects} &\makecell{Visual\\Inspection}	&ResGNet-C	&--	&--	&5	&\makecell{Acc=96.62\\Sen=97.33\\Spec=95.91\\Prec=96.21}\\
\hline
\cite{a6} & \makecell{COVID-CT\\Dataset} & CT & \makecell{349 COVID-19 Images,\\397 Non-COVID-19 Images} & \makecell{Scaling Process,\\DA} & \makecell{Multiple\\Kernels-ELM\\-based DNN} & -- & Matlab & 10 & \makecell{Acc=98.36\\Sen=98.28\\Spec=98.44\\Prec=98.22}\\
\hline
\cite{a7} & Clinical & CT & \makecell{210,395 Images From 704\\ COVID-19 Patients and\\498 Non-COVID-19 Subjects} & DA & \makecell{U-net\\ \hline Dual-Branch\\Combination\\Network} & Attention Maps& PyTorch & 5 &\makecell{Acc=92.87\\Sen=92.86\\Spec=92.91}\\
\hline
\cite{a8} & Various Dataset & CT & \makecell{2933 COVID-19 Images} & \makecell{Deleting Outliers,\\Normalization,\\Resizing} & \makecell{Ensemble\\DNN} & -- & Matlab R2019a & 5 & \makecell{Acc= 99.054\\Sen= 99.05\\Spec=99.6\\F1-Score= 98.59}\\
\hline
\cite{a9} & Clinical & CT & \makecell{320 COVID-19 Images,\\320 Healthy Control Images} & \makecell{Histogram\\Stretching,\\Margin Crop,\\Resizing,\\Down Sampling} & FGCNet & \makecell{Gradient-\\Weighted CAM\\ (Grad-CAM)} & -- &-- &\makecell{Acc=97.14\\Sen=97.71\\Spec=96.56\\Prec=96.61}\\
\hline
\cite{a10}	&Clinical	&CT	&\makecell{180 Viral Pneumonia,\\94 COVID-19 Cases}	&\makecell{ROIs\\Extraction}	&\makecell{Modified\\Inception}	&--	&--	&--	&\makecell{Acc=89.5\\Sen=88\\Spec=87\\F1-Score=77}\\

\hline
\cite{a11}	&Clinical	&CT	&\makecell{3389 COVID-19 Images,\\1593 Non-COVID-19 Images} &\makecell{Segmentation,\\ Generating\\Lung Masks}	& \makecell{3D ResNet34\\with\\Online\\Attention}	&Grad-CAM	&PyTorch	&5	&\makecell{Acc=87.5\\Sen=86.9\\Spec=90.1\\F1-Score=82.0}\\
\hline
\cite{a12}	&\makecell{COVIDx-CT\\Dataset}	&CT	&\makecell{104,009 Images From\\1,489 Patient Cases}	&\makecell{Automatic\\Cropping\\Algorithm, DA}	&COVIDNet-CT	&--	&TensorFlow	&--	&\makecell{Acc= 99.1\\Sen=97.3\\PPV=99.7}\\
\hline
\cite{a13}	&Various Datasets	&CT	&\makecell{349 COVID-19 Images,\\397 Non-COVID-19 Images}	& \makecell{Resizing,\\Normalization,\\Wavelet-Based\\DA}	&ResNet18	&\makecell{Localization of\\Abnormality}	&Matlab 2019b	&--	&\makecell{Acc=99.4\\Sen=100\\Spec=98.6}\\
\hline
\cite{a14}	&COVID-CT &CT	&\makecell{345 COVID-19 Images,\\397 Non-COVID-19 Images}	&Resizing, DA	&\makecell{Conditional\\GAN\\ \hline ResNet50}	&--	&\makecell{TensorFlow,\\Matlab}	&--	&\makecell{Acc=82.91\\Sen=77.66\\Spec=87.62}\\
\hline
\cite{a15}	&Clinical	&CT	&\makecell{151 COVID-19 Patient,\\498 Non-COVID-19 Patient}	&\makecell{Resizing,\\Padding, DA}	&3D-CNN	&\makecell{Interpretation\\by Two\\Radiologists}	&--	&--	&AUC=70\\
\hline
\cite{a16} & \makecell{SARS-CoV-2\\CT-Scan Dataset} & CT & \makecell{1252 CT COVID-19 Images,\\1230 CT non-COVID-19 Images} & -- & \makecell{GAN with\\Whale\\Optimization\\Algorithm} & -- & Matlab 2020a & 10 & \makecell{Acc=99.22\\Sen=99.78\\Spec=97.78\\F1-score=98.79}\\
\hline
\cite{a17}	&Various Datasets	&CT	&\makecell{1,684 COVID-19 Patient,\\1,055 Pneumonia,\\914 Normal Patients}	&Resizing 	&Inception V1	& \makecell{Interpretation by\\6 Radiologists,\\t-SNE Method}	&--	&10	&\makecell{Acc=95.78\\AUC=99.4}\\
\hline
\cite{a18}	&Clinical	&CT	&\makecell{2267 COVID-19 CT Images,\\1235 HC CT Images} 	& \makecell{Compressing,\\Normalization,\\Cropping,\\Resizing}	&ResNet50	&--	&Keras	&--	&\makecell{Acc=93\\Sen=93\\Spec=92\\F1-Score=92}\\
\hline
\cite{a19}	&Clinical	&CT	&\makecell{108 COVID-19 Patients,\\86 Non-COVID-19 Patients}	&\makecell{Visual\\Inspection,\\Grey-Scaling,\\Resizing}	&\makecell{Various\\Networks}	&--	&--	&--	&\makecell{Acc=99.51\\Sen=100\\Spec=99.02}\\
\hline
\cite{a20}&	Various Datasets	&CT	&\makecell{413 COVID-19 Images,\\439 Non-COVID-19 Images}	& \makecell{Feature Extraction\\with ResNet-50}	& 3D-CNN	&-- 	&--	&10	&\makecell{Acc=93.01\\Sen=91.45\\Spec=94.77\\Prec=94.77}\\
\hline

\cite{a21}	&Clinical	&CT	&\makecell{150 3D COVID-19 Chest CT,\\CAP and NP Patients\\(450 Patient Scans)}	&\makecell{Sliding\\Window, DA} &\makecell{Multi-View\\U-Net\\ \hline 3D-CNN}	& \makecell{Weakly\\Supervised\\Lesions\\Localization,\\CAM}	&TensorFlow	&5	&\makecell{Acc=90.6\\Sen=83.3\\Spec=95.6\\Prec=74.1}\\
\hline
\cite{a22}	&Various Datasets	&CT	&\makecell{449 COVID-19 Patients,\\425 Normal, 98 Lung Cancer,\\397 Different Kinds of\\Pathology} &\makecell{Resizing,\\Intensity\\Normalization} &	\makecell{Autoencoder\\Based\\DNN}	&--	&\makecell{Keras,\\TensorFlow}	&--	&\makecell{Dice=88\\Acc=94.67\\Sen=96\\Spec=92}\\
\hline
\cite{a23}	&\makecell{COVID-19 CT\\from \cite{a17}}	&CT	&746 Images	&--	&GAN	&--	&Matlab	&--	&\makecell{Acc=84.9\\Sen=85.33\\Prec=85.33}\\
\hline
\cite{a24}	& \makecell{COVID-19 CT\\Datasets, Cohen}	&CT	&\makecell{345 COVID-19 CT Images,\\375 Non-COVID-19 CT Image}	&\makecell{2D Redundant\\Discrete WT\\(RDWT) Method,\\Resizing}	&ResNet50	&\makecell{Grad-CAM,\\Occlusion\\Sensitivity\\Technique}	&Matlab	&10	&\makecell{Acc=92.2\\Sen=90.4\\Spec=93.3\\F1-Score=91.5}\\

\hline
\cite{a25}	& \makecell{SARS-CoV-2\\CT Scan Dataset}	&CT	&\makecell{1262 COVID-19 Images,\\1230 HC Images} 	&--	&\makecell{Convolutional\\Support Vector\\Machine\\(CSVM)}	&--	&Matlab 2020a	&--	&\makecell{Acc=94.03\\Sen=96.09\\Spec=92.01\\Pre=92.19}\\
\hline
\cite{a26}	&\makecell{Chest CT\\and X-ray}	&\makecell{X-Ray,\\CT} &\makecell{5857 Chest X-Rays,\\767 Chest CTs}	&--	&\makecell{Various\\Networks}	&Heat Map	&Keras	&--	&\makecell{Acc=75\\(CT)}\\
\hline
\cite{a27} & \makecell{medseg\\DlinRadiology} & CT & \makecell{10 Axial Volumetric CTS\\(Each Containing 100 Slices\\of COVID-19 Images)} & Resizing & \makecell{VGG16,\\Resnet-50\\ \hline U-net} & -- & -- & -- & \makecell{Acc=99.4\\Spec=99.5\\Sen=80.83\\Dice=72.4\\IOU=61.59}\\
\hline
\cite{a28}	&BasrahDataset	&CT 	&50 Cases, 1425 Images	&\makecell{Gray-Scaling,\\Resizing}	&VGG 16	&--	&Keras	&--	&\makecell{Acc=99\\F1-Score=99}\\
\hline
\cite{a29}	&Kaggle	&CT 	&\makecell{1252 COVID CT Images,\\1240 non-COVID CT Images}	&\makecell{Resizing,\\Normalization,\\DA}	&Covid CT-net	&heat map	&\makecell{TensorFlow,\\Keras}	&--	&\makecell{Acc=95.78\\Sen=96\\Spec=95.56}\\
\hline
\cite{a30}	&COVID-CT	&CT 	&\makecell{708 CTs, 312 with COVID-19,\\396 Non-COVID-19}	&Normalization	&LeNet-5	&--	&--	&5	&\makecell{Acc=95.07\\Sen=95.09\\Prec=94.99}\\
\hline
    
    \end{longtable}
\end{center}

\section{Materials and Methods}

This section of the paper is devoted to discussing the applied method and its components. In this paper, we have firstly collected a new CT scan dataset from COVID-19 patients; then, from each scan, the informative slices were selected by physicians. After that, several convolutional neural networks pre-trained on the ImageNet dataset \cite{r1} were fine-tuned to the task at hand. Here, we trained a Resnet-50 architecture \cite{r2}, an EfficientNet B3 architecture \cite{r3}, a Densenet-121 architecture \cite{r4}, a ResNest-50 architecture \cite{r5}, and a ViT architecture \cite{r6}. Several data augmentation techniques alongside a CycleGAN model were applied to improve the performance of each network further.

Here, the details of each step are presented; firstly, an explanation is given on the applied dataset. Specifications of each applied deep neural network (DNN) are discussed afterward. Finally, CycleGAN is explained in the last part alongside the overall proposed method.

\subsection{Dataset}

\begin{figure}[h]
\centering

\begin{subfigure}
    \centering
    \includegraphics[width=0.17\textwidth]{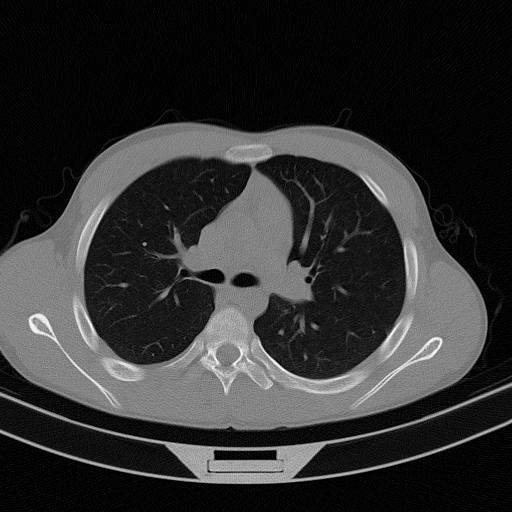}
\end{subfigure}
\hspace{2pt}
\begin{subfigure}
    \centering
    \includegraphics[width=0.17\textwidth]{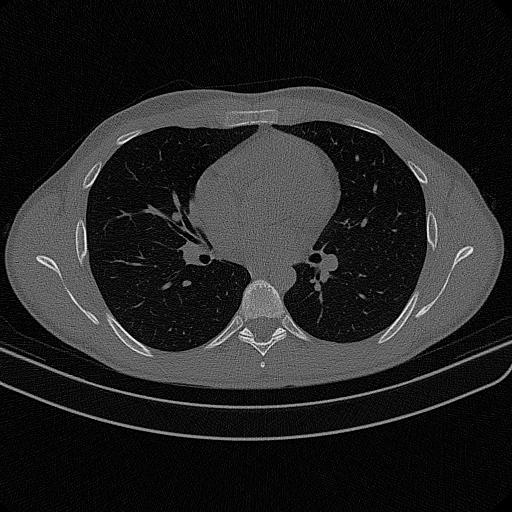}
\end{subfigure}
\hspace{2pt}
\begin{subfigure}
    \centering
    \includegraphics[width=0.17\textwidth]{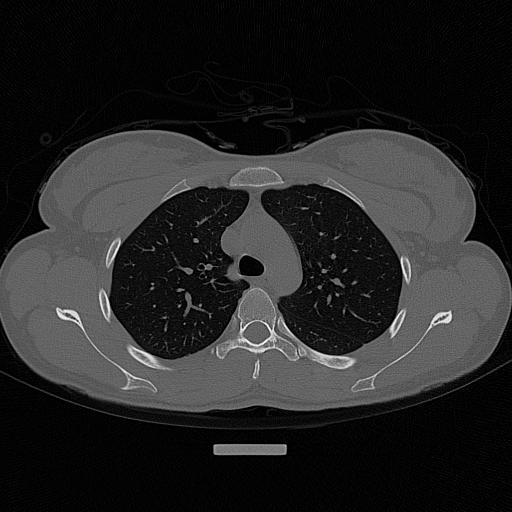}
\end{subfigure}
\hspace{2pt}
\begin{subfigure}
    \centering
    \includegraphics[width=0.17\textwidth]{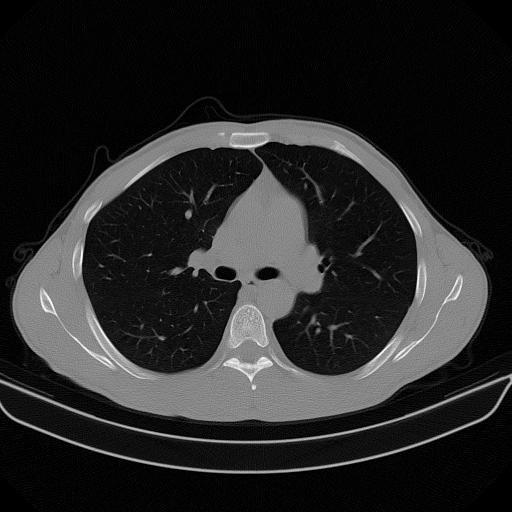}
\end{subfigure}
\hspace{2pt}
\begin{subfigure}
    \centering
    \includegraphics[width=0.17\textwidth]{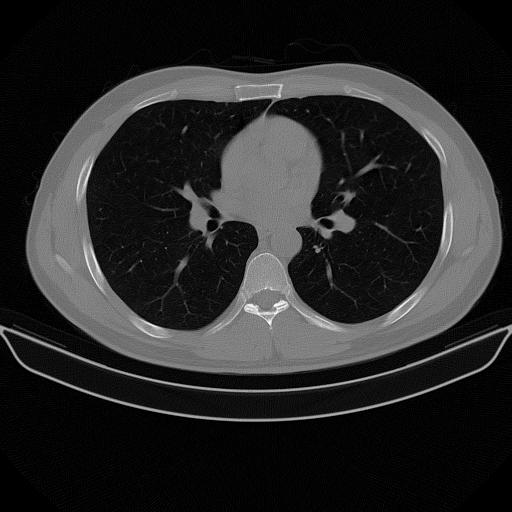}
\end{subfigure}

\begin{subfigure}
    \centering
    \includegraphics[width=0.17\textwidth]{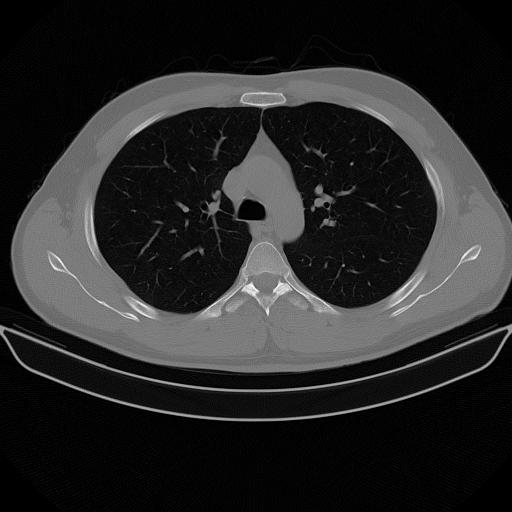}
\end{subfigure}
\hspace{2pt}
\begin{subfigure}
    \centering
    \includegraphics[width=0.17\textwidth]{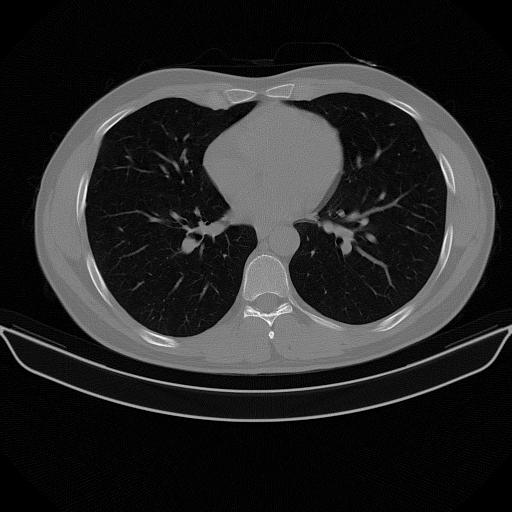}
\end{subfigure}
\hspace{2pt}
\begin{subfigure}
    \centering
    \includegraphics[width=0.17\textwidth]{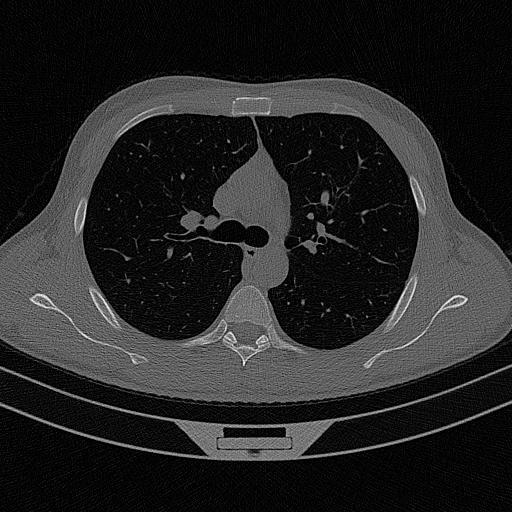}
\end{subfigure}
\hspace{2pt}
\begin{subfigure}
    \centering
    \includegraphics[width=0.17\textwidth]{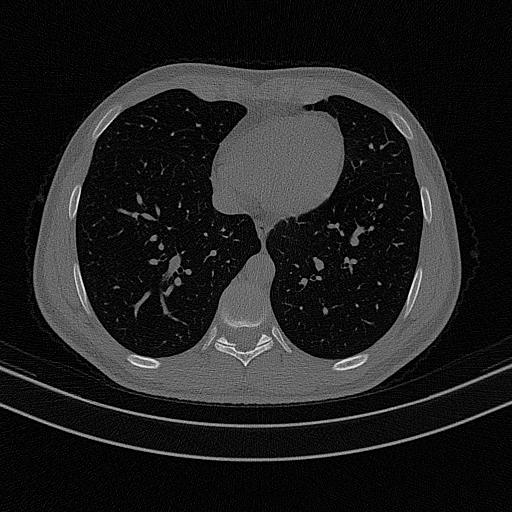}
\end{subfigure}
\hspace{2pt}
\begin{subfigure}
    \centering
    \includegraphics[width=0.17\textwidth]{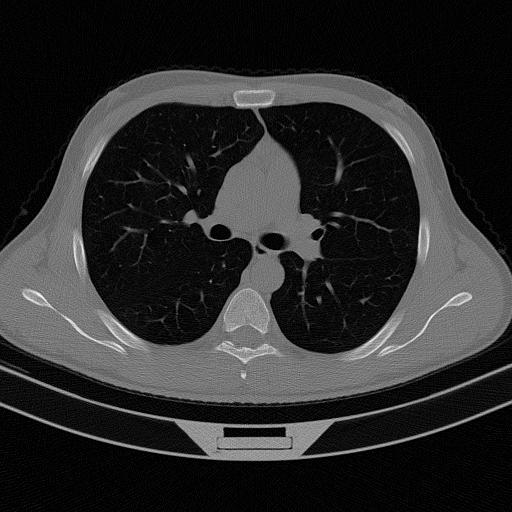}
\end{subfigure}

\begin{subfigure}
    \centering
    \includegraphics[width=0.17\textwidth]{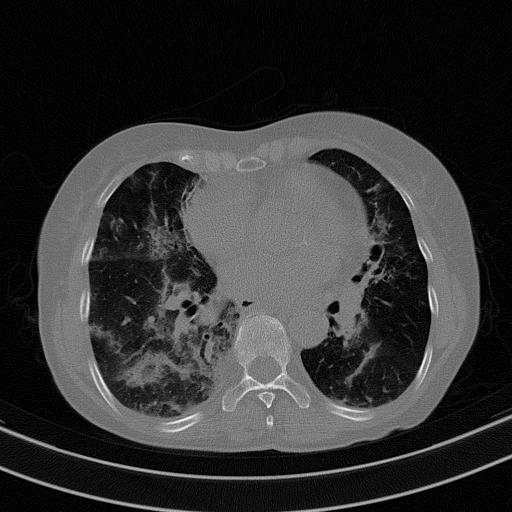}
\end{subfigure}
\hspace{2pt}
\begin{subfigure}
    \centering
    \includegraphics[width=0.17\textwidth]{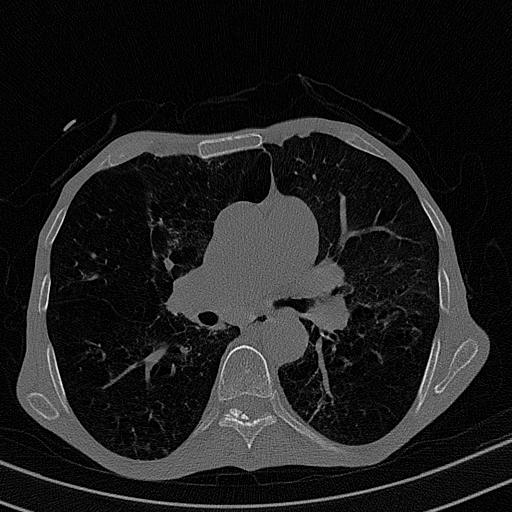}
\end{subfigure}
\hspace{2pt}
\begin{subfigure}
    \centering
    \includegraphics[width=0.17\textwidth]{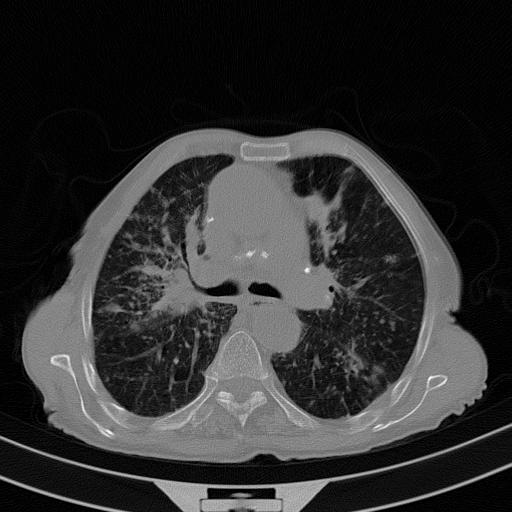}
\end{subfigure}
\hspace{2pt}
\begin{subfigure}
    \centering
    \includegraphics[width=0.17\textwidth]{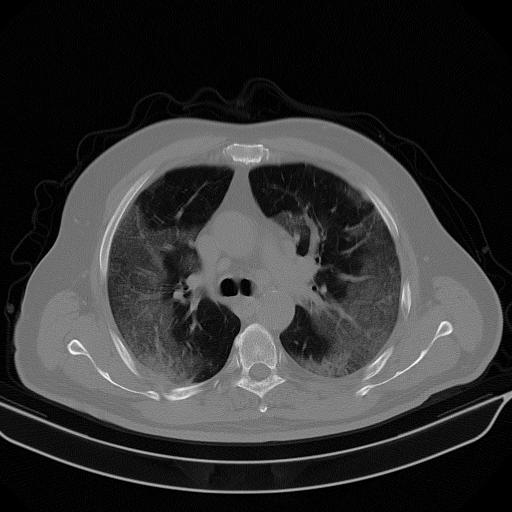}
\end{subfigure}
\hspace{2pt}
\begin{subfigure}
    \centering
    \includegraphics[width=0.17\textwidth]{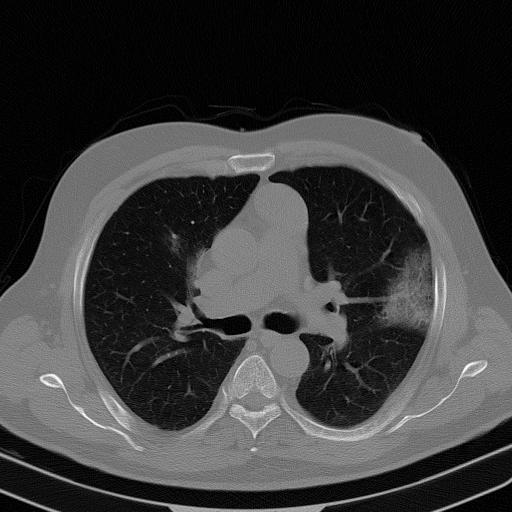}
\end{subfigure}

\begin{subfigure}
    \centering
    \includegraphics[width=0.17\textwidth]{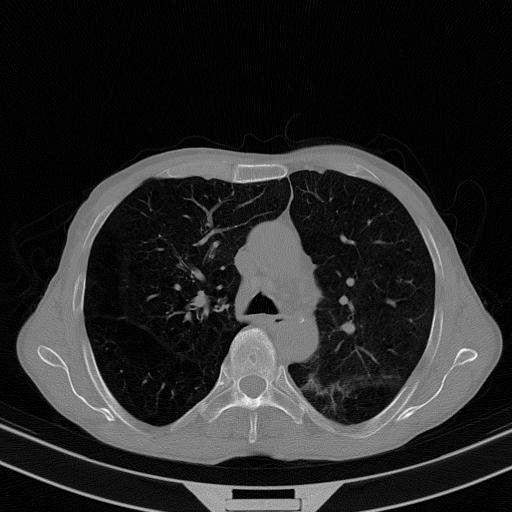}
\end{subfigure}
\hspace{2pt}
\begin{subfigure}
    \centering
    \includegraphics[width=0.17\textwidth]{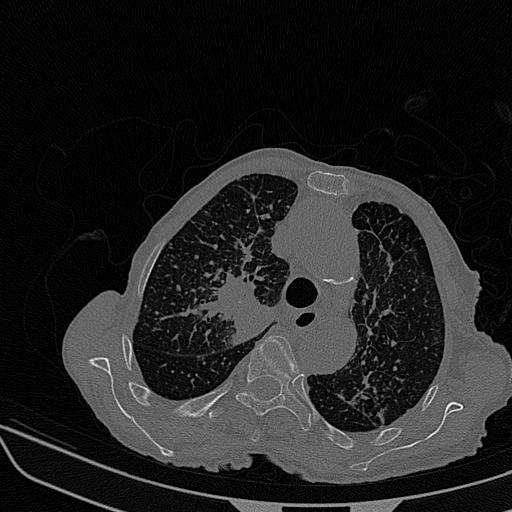}
\end{subfigure}
\hspace{2pt}
\begin{subfigure}
    \centering
    \includegraphics[width=0.17\textwidth]{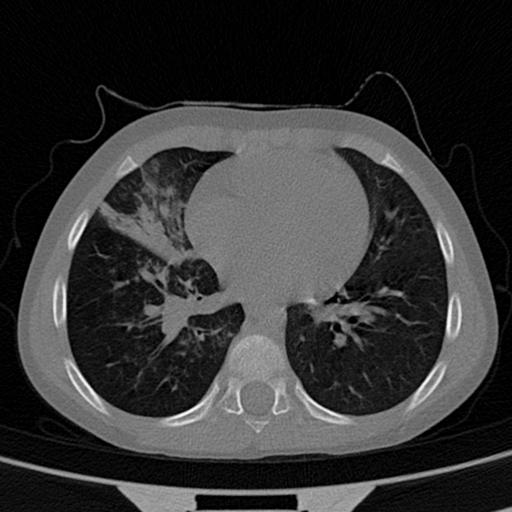}
\end{subfigure}
\hspace{2pt}
\begin{subfigure}
    \centering
    \includegraphics[width=0.17\textwidth]{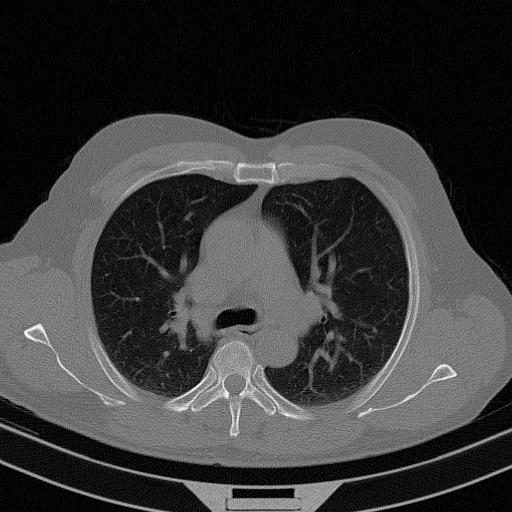}
\end{subfigure}
\hspace{2pt}
\begin{subfigure}
    \centering
    \includegraphics[width=0.17\textwidth]{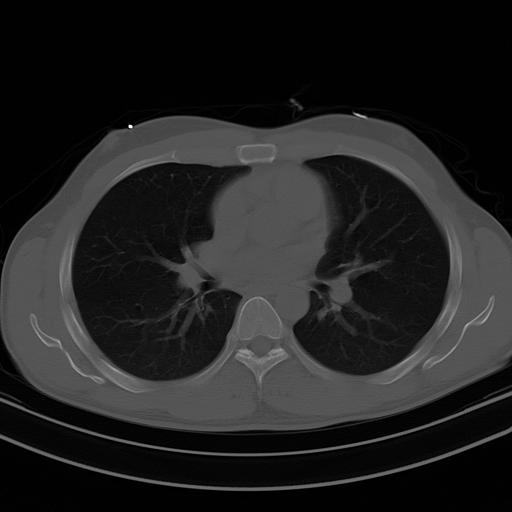}
\end{subfigure}

\caption{Examples of dataset, first two rows contain images from healthy subjects, whereas the last two rows contain images from COVID-19 patients.\label{figdata}}

\end{figure}

In this paper, a new CT scans dataset of COVID-19 patients was collected from Gonabad Hospital in Iran; all data were recorded by radiologists between June 2020 and December 2020. The number of subjects with COVID-19 is 90, and 99 of the subjects are normal. It is noteworthy to mention that the normal subjects are patients with suspicious symptoms and not merely a control group; this makes this dataset unique compared to its prior ones, as they usually have used scans of other diseases for the control group. Patients with COVID-19 or normal subjects range in age from 2 to 88 years; 69 of which are female and 120 are male (both COVID-19 and normal classes). A total of 1766 slices of these scans were finally selected by specialist physicians from the abnormal class and 1397 from the normal class; the labeling of each CT image was performed by three experienced radiologists along with two infectious disease physicians. In addition, RT-PCR was taken from each subject to confirm labelings. All ethical approvals have been obtained from the hospital to use CT scans of COVID-19 patients and normal individuals for research purposes. Figure \ref{figdata} illustrates a few CT scans of healthy individuals and patients with COVID-19.

\subsection{Deep Neural Networks}

\subsubsection{Resnet}
ResNet \cite{r2} architecture was introduced in 2015 with a depth of 152 layers; it is known as the deepest architecture up to that year and still is considered as one of the deepest. There are various versions of the architecture with different depths that are used depending on the need. This network's main idea was to use a block called residual block, which tried to solve the problem of vanishing gradients, allowing the network to go deeper without reducing performance. Proving its capabilities by winning the Image Net Challenge in 2015; the ideas of this network have been applied in many others ever since. In this paper, a version of this network with a depth of 50 has been used, which is a wise choice given the considerably smaller amount of data compared to the ImageNet database.

\subsubsection{EfficientNet}

Three different criteria must be tested to design a convolutional neural network: the depth, width, and resolution of the input images. Choosing the proper values of the three criteria in such a way that they form a suitable network together is a challenging task. Increasing the depth can lead to finding complex patterns, but it can also cause problems such as vanishing gradients. More width can increase the quality of the features learned, but accuracy for such network tends to quickly saturate. Also, high image quality can have a detrimental effect on accuracy. The network was introduced in \cite{r3} with a study on how to scale the network in all three criteria properly. Using a step-by-step scheme, the network first finds the best structure for a small dataset and then scales that structure according to the activity. The network has been used for many tasks, including diagnosing autism \cite{no} and schizophrenia \cite{ns}.

\subsubsection{Densenet}

Introduced by Huang et al. \cite{r4}, DenseNet, densely connected convolutional networks, has improved the baseline performance on benchmark computer vision task and shown its efficiency. Utilizing residuals in a better approach has allowed this network to exploit fewer parameters and go deeper. Also, by feature reuse, the number of parameters is reduced dramatically. Its building blocks are dense blocks and transition layers. Compared to ResNet, DenseNet uses concatenation in residuals rather than summing them up. To make this possible, each feature vector of each layer is chosen to have the same size for each dense block; also, training these networks has been shown to be easier than prior ones \cite{r4}. This is arguably due to the implicit deep supervision where the gradient is flowing back more quickly. The capability to have thin layers is another remarkable difference in DenseNet compared to other state-of-the-art techniques. The parameter K, the growth rate, determines the number of features for each layer's dense block. These feature vectors are then concatenated with the preceding ones and given as input to the subsequent layer. Eliminating optimization difficulties for scaling up to hundreds of layers is another DenseNet superiority.

\subsubsection{ViT}

Arguably, the main problem with convolutional neural networks is their failure in encoding relative spatial information. In order to overcome this issue, researchers in \cite{r6} have adopted the self-attention mechanism from natural language processing (NLP) models. Basically, attention can be defined as trainable weights that model each part of an input sentence's importance. Changing networks from NLP to computer vision, pixels are picked as parts of the image to train the attention model on them. Nevertheless, pixels are very small parts of an image; thus, one can pick a bigger segment of an image as one of its parts, i.e., a 16 by 16 block of images. ViT uses a similar idea; by dividing the image into smaller patches to train the attention model on them. Also, ViT-Large has 24 layers with a hidden size of 1,024 and 16 attention heads. Examination shows not only superior results but also significantly reduced training time and also less demand for hardware resources \cite{r6}.

\subsubsection{ResNest}

Developed by researchers from Amazon and UC Davis, ResNest \cite{r5} is also another attention-based neural network that has also adopted the ideas behind ResNet structure. In its first appearance, this network has shown significant performance improvement without a large increase in the number of parameters, surpassing prior adaptations of ResNet such as ResNeXt and SEnet. In their paper, they have proposed a modular Split-Attention block that can distribute attention to several feature-map groups. The split-attention block is made of the feature-map group and split-attention operations; then, by stacking those split-attention blocks similar to ResNet, researchers were able to produce this new variant. The novelties of their paper are not merely introducing a new structure, but they also introduced a number of training strategies.

\subsection{Data Augmentation and Training Process}

Generative adversarial networks were first introduced in 2014 \cite{ngan} and found their way into various fields shortly after \cite{ndeep}. They have also been used as a method for data augmentation, and network pretraining \cite{nngan} previously as well. A particular type of these networks is CycleGAN \cite{r10}, a network created mainly for unpaired image-to-image translation. In this particular form of image-to-image translation, there is no need for a dataset containing paired images, which is itself a challenging task. The CycleGAN comprises of training of two generator discriminators simultaneously. One generator uses the first group of images as input and creates data for the second group, and the other generator does the opposite. Discriminator models are then utilized to distinguish the generated data from real ones and feed the gradients to generators subsequently.

\begin{figure}[h]	

\includegraphics[width=\textwidth]{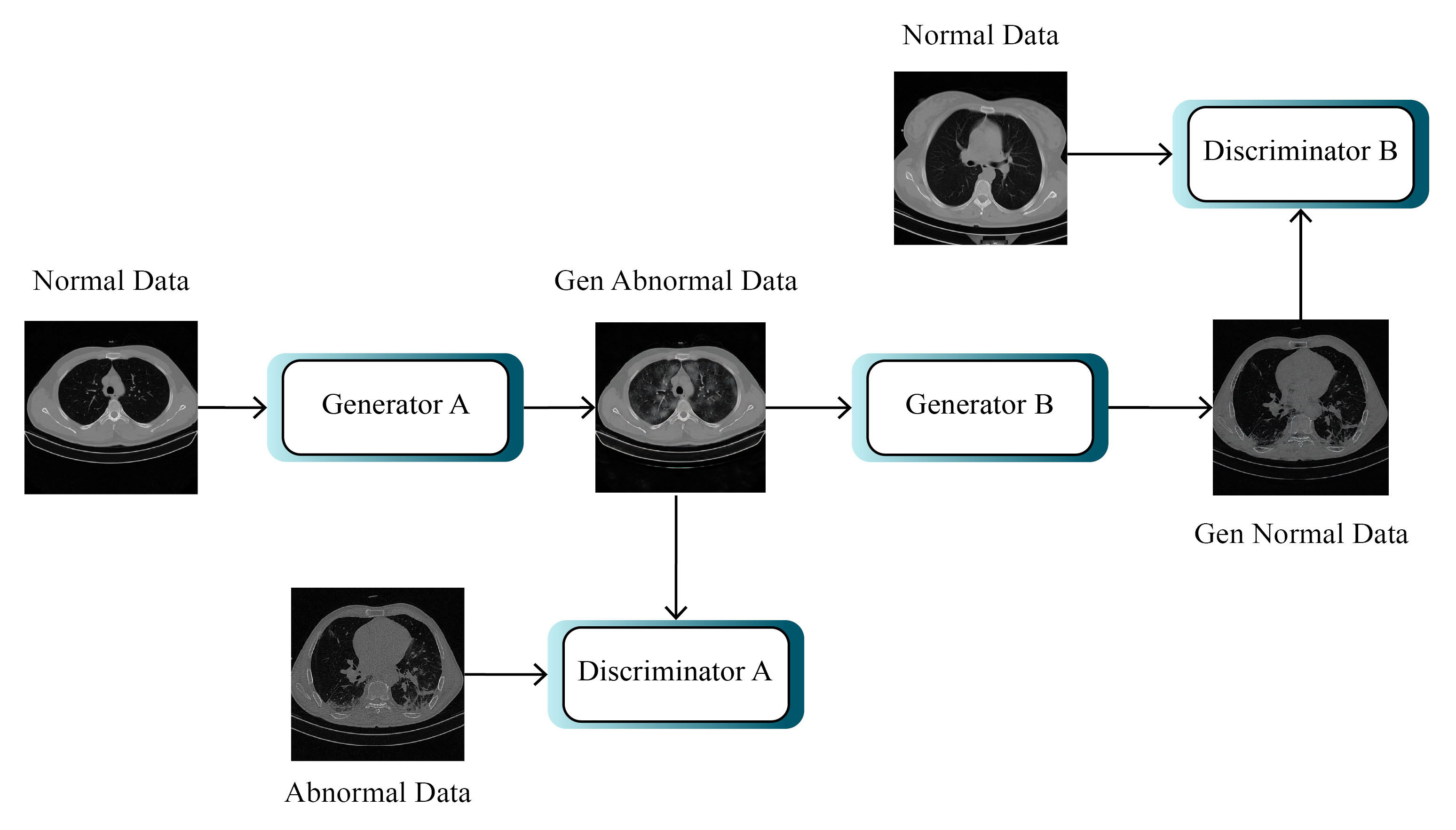}
\caption{Overall diagram of applied CycleGAN.\label{figtwo}}
\end{figure}

\begin{figure}[!ht]
\centering
\begin{subfigure}
    \centering
    \includegraphics[width=0.2\textwidth]{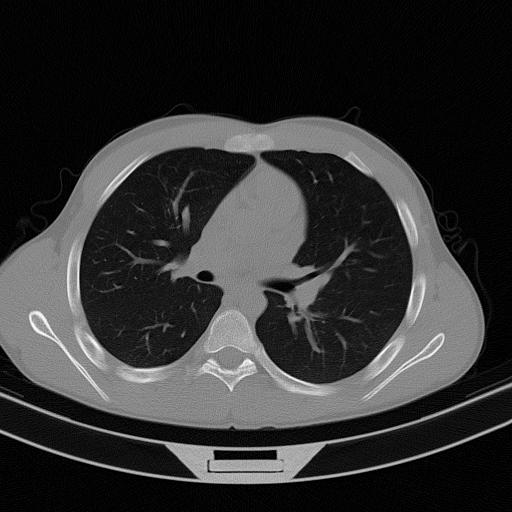}
\end{subfigure}
\hspace{2pt}
\begin{subfigure}
    \centering
    \includegraphics[width=0.2\textwidth]{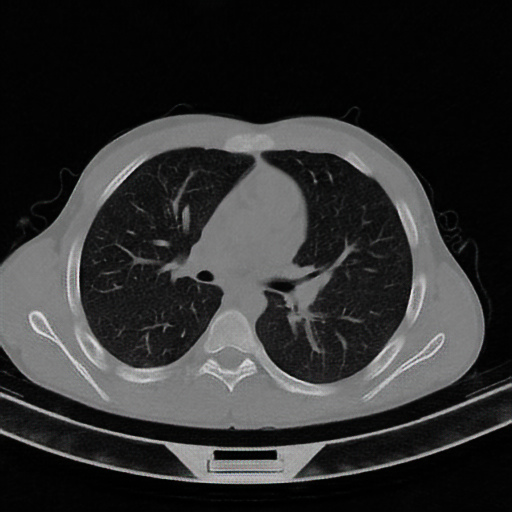}
\end{subfigure}
\hspace{2pt}
\begin{subfigure}
    \centering
    \includegraphics[width=0.2\textwidth]{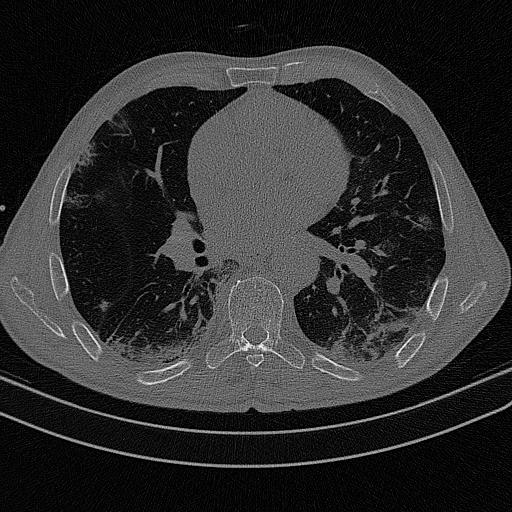}
\end{subfigure}
\hspace{2pt}
\begin{subfigure}
    \centering
    \includegraphics[width=0.2\textwidth]{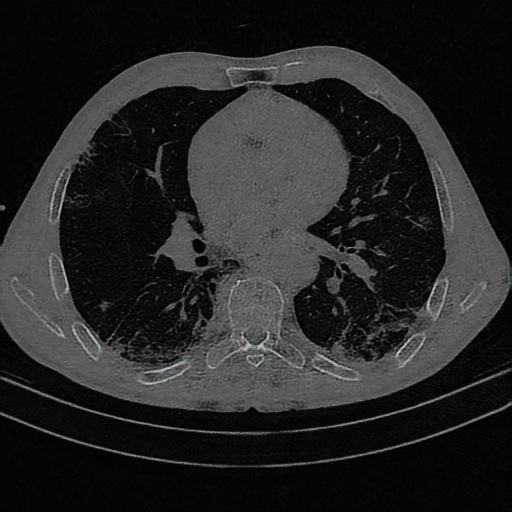}
\end{subfigure}

\begin{subfigure}
    \centering
    \includegraphics[width=0.2\textwidth]{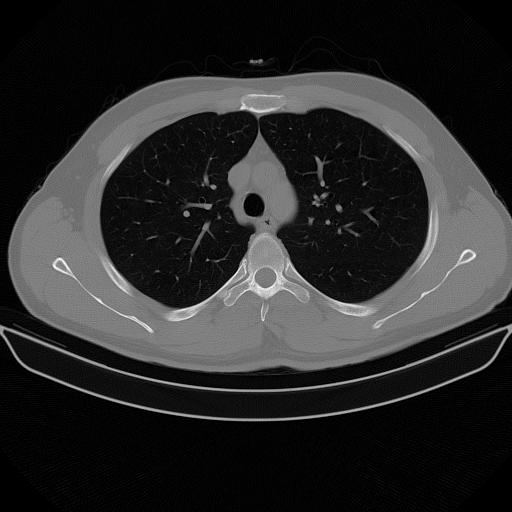}
\end{subfigure}
\hspace{2pt}
\begin{subfigure}
    \centering
    \includegraphics[width=0.2\textwidth]{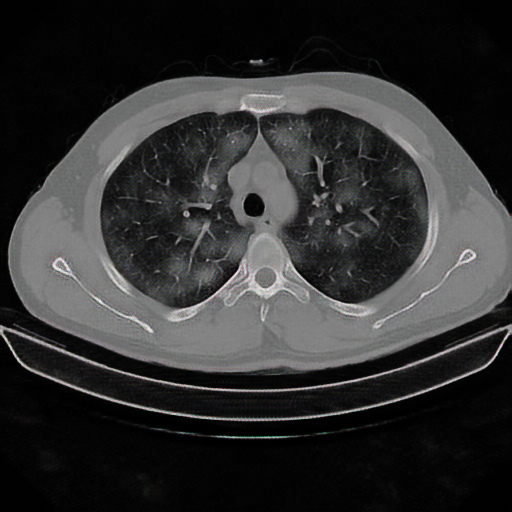}
\end{subfigure}
\hspace{2pt}
\begin{subfigure}
    \centering
    \includegraphics[width=0.2\textwidth]{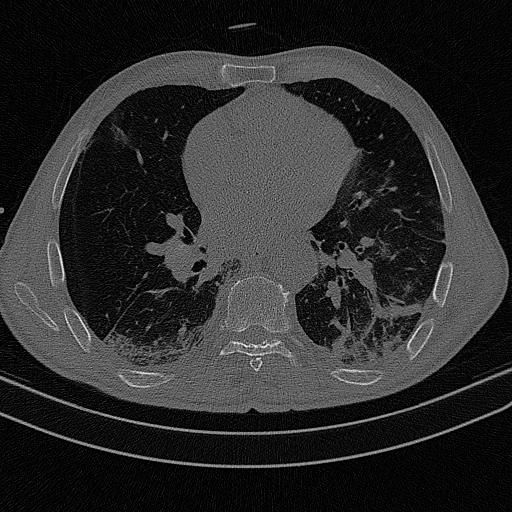}
\end{subfigure}
\hspace{2pt}
\begin{subfigure}
    \centering
    \includegraphics[width=0.2\textwidth]{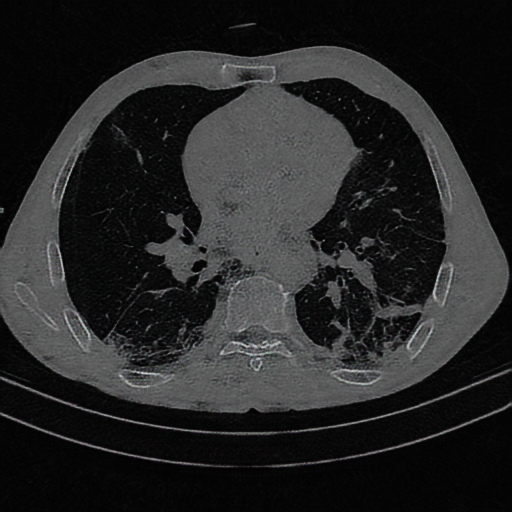}
\end{subfigure}

\begin{subfigure}
    \centering
    \includegraphics[width=0.2\textwidth]{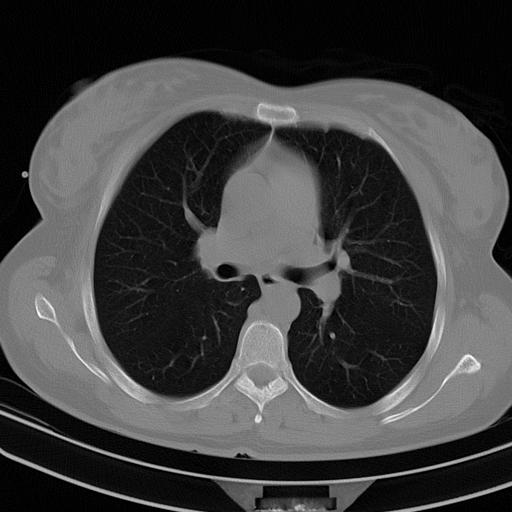}
\end{subfigure}
\hspace{2pt}
\begin{subfigure}
    \centering
    \includegraphics[width=0.2\textwidth]{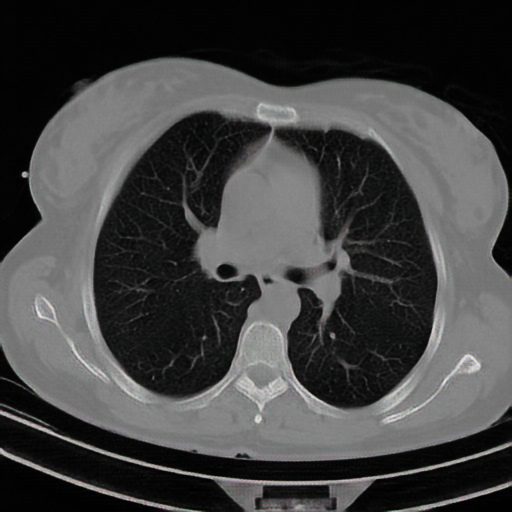}
\end{subfigure}
\hspace{2pt}
\begin{subfigure}
    \centering
    \includegraphics[width=0.2\textwidth]{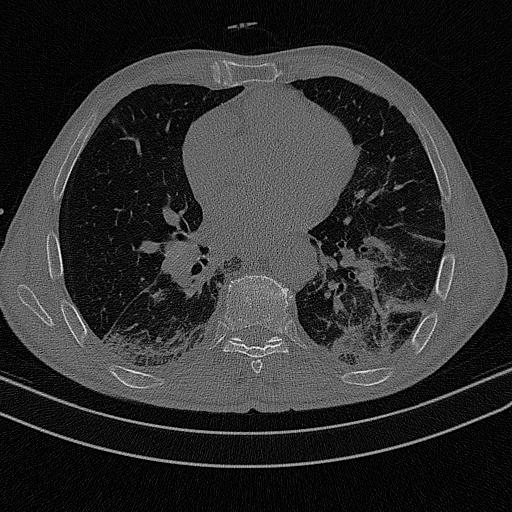}
\end{subfigure}
\hspace{2pt}
\begin{subfigure}
    \centering
    \includegraphics[width=0.2\textwidth]{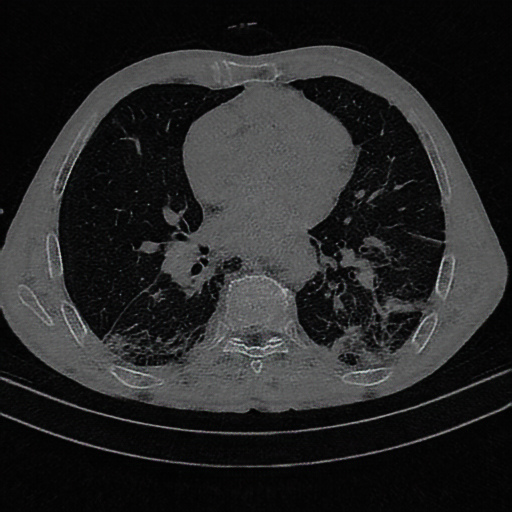}
\end{subfigure}

\begin{subfigure}
    \centering
    \includegraphics[width=0.2\textwidth]{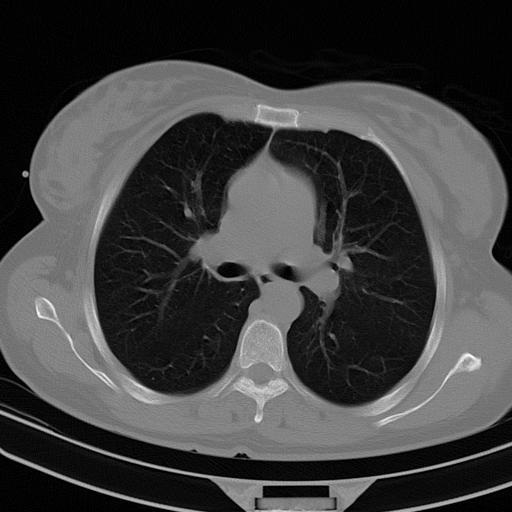}
\end{subfigure}
\hspace{2pt}
\begin{subfigure}
    \centering
    \includegraphics[width=0.2\textwidth]{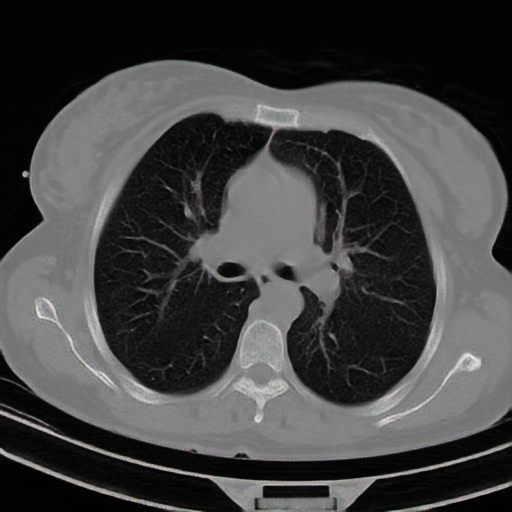}
\end{subfigure}
\hspace{2pt}
\begin{subfigure}
    \centering
    \includegraphics[width=0.2\textwidth]{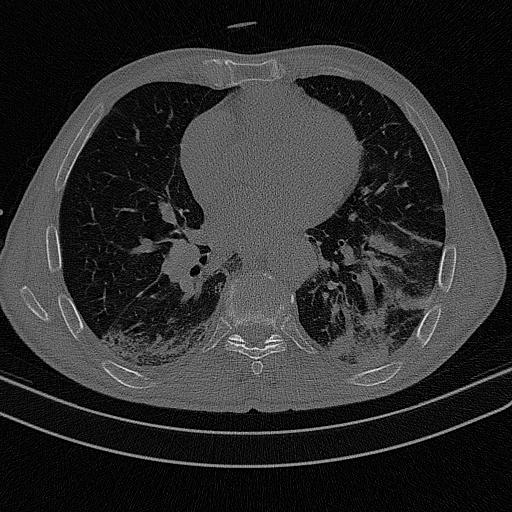}
\end{subfigure}
\hspace{2pt}
\begin{subfigure}
    \centering
    \includegraphics[width=0.2\textwidth]{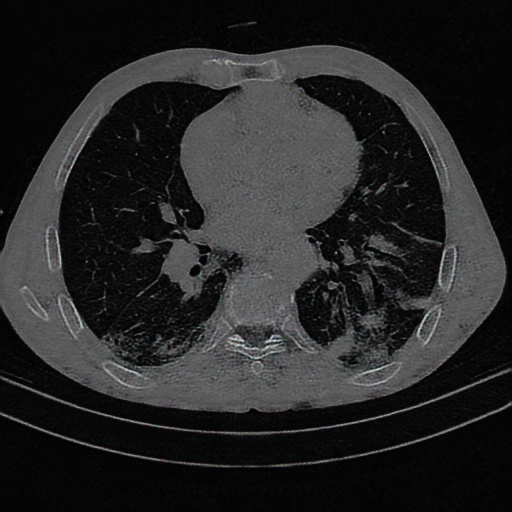}
\end{subfigure}

\begin{subfigure}
    \centering
    \includegraphics[width=0.2\textwidth]{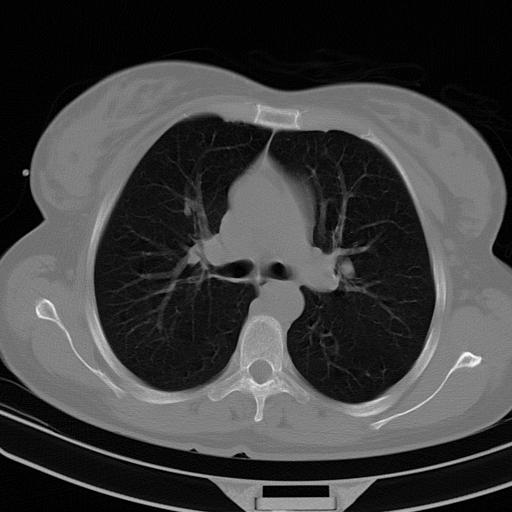}
\end{subfigure}
\hspace{2pt}
\begin{subfigure}
    \centering
    \includegraphics[width=0.2\textwidth]{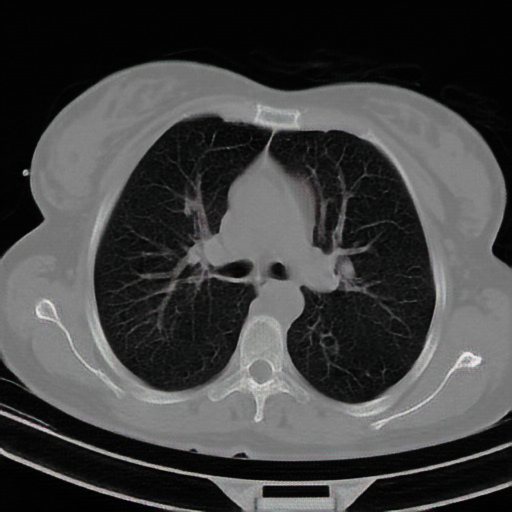}
\end{subfigure}
\hspace{2pt}
\begin{subfigure}
    \centering
    \includegraphics[width=0.2\textwidth]{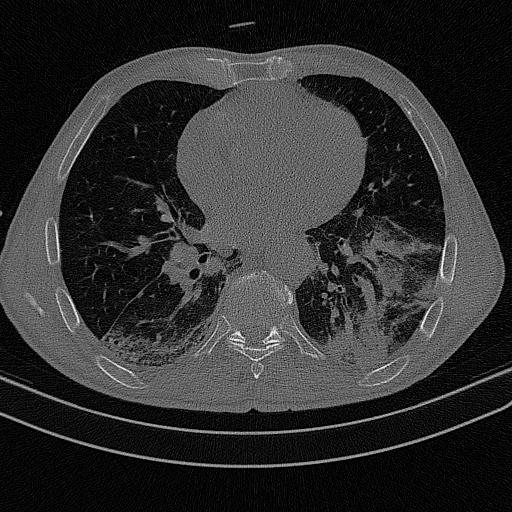}
\end{subfigure}
\hspace{2pt}
\begin{subfigure}
    \centering
    \includegraphics[width=0.2\textwidth]{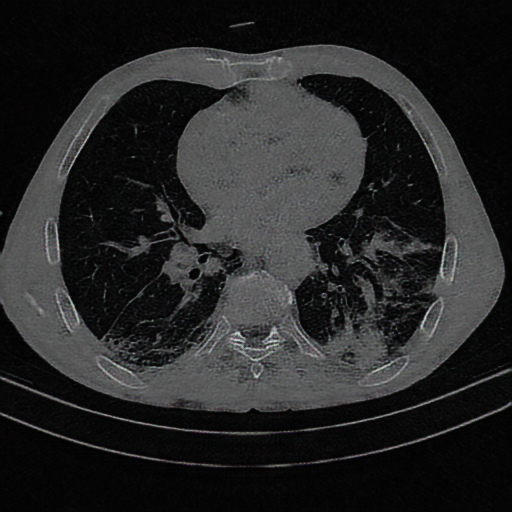}
\end{subfigure}

\caption{Examples of generated data, the first column shows normal data from the main dataset. The second column shows the generated abnormal data from those images. The third column shows abnormal data from the main dataset. Lastly, the fourth column shows the generated normal data from those images.\label{figgen}}

\end{figure}

The CycleGAN used in this paper has a similar structure to the one presented in the main paper \cite{r10}. Compared to other GAN paradigms, CycleGAN uses image-to-image translation, which simplifies the training process, especially where training data is limited, which also helps to create data of the desired class easily. However, using other GAN paradigms, such as conditional GAN \cite{nnfive}, one can also create data of a specific class, yet training those methods is more complicated. A diagram of the CycleGAN is presented in Figure \ref{figtwo}, and also a few samples of generated data are illustrated in Figure \ref{figgen}.

In this paper, to train the networks properly, first, we preprocessed images by applying a Gaussian filter. Then, we applied several data augmentation techniques \cite{r9}, namely, by using random flips, rotations, zooms, warps, lighting transforms, and also presizing \cite{r7}. We also studied our models' performance by training them using an augmented dataset generated by means of a CycleGAN model implemented using the UPIT library \cite{nnfour}. 

\section{Results}

\subsection{Environment Setup and Hyper Parameter Selection}

All models were trained using the FastAI library \cite{r7} and applying fine-tuning to the pre-trained models available at the timm repository \cite{nnthree} using a GPU Nvidia RTX 2080 Ti with 11 GB of RAM. As for the CycleGAN implementation, the UPIT library \cite{nnfour} was used. To find the best hyperparameters, such as the learning rate for the task at hand, and to evaluate our models properly, we divided the data into three parts: the first one for training, the second one for validation, and the last one for testing. This division was done using a 70/15/15 scheme, and also, no two slices of any patient are presented in two different parts simultaneously to make the results trustworthy. To set the learning rate for the architectures, we employed the two-stage procedure similar to the one presented in \cite{r3}; lastly, we applied early stopping in all the architectures to avoid overfitting. The final selected values for batch size and hyperparameters are all available in Table \ref{tabhyper}.

\begin{table}[h] 
\caption{Selected hyperparameters for each network.\label{tabhyper}}
\centering
\resizebox{0.6\textwidth}{!}{
\begin{tabular}{|c|c|c|}
\hline
\textbf{Network}	& \textbf{Batch Size}	& \textbf{Learning Rate}\\
\hline
Densenet-121	&16	&1.00E-03\\
\hline
EfficientNet-B3	&16	&1.00E-03\\
\hline
Resnet-50	&16	&1.00E-03\\
\hline
ResNeSt-50	&16	&1.00E-04\\
\hline
ViT	&16	&1.00E-05\\
\hline
\end{tabular}}
\end{table}

\subsection{Evaluation Metrics}

The evaluation of each network's performance is measured by several different statistical metrics, considering that merely relying on one measure of accuracy, it is not possible to measure all the different aspects of the performance of a network. The metrics used in this article are accuracy, precision, recall, F1-score, and area under receiver operating characteristic (ROC) curve (AUC) \cite{n1}. How to calculate these metrics is also shown in Table 2. In this table, TP shows the number of positive cases that have been correctly classified, TN has shown the number of negative cases that have been correctly classified, and FP and FN are the numbers of positive and negative cases that have been misclassified, respectively. In addition, for each network, a learning curve is plotted that shows the speed of learning and how to converge.

\begin{table}[h] 
\caption{Statistical metrics for performance evaluation.\label{tabmet}}
\centering
\resizebox{0.6\textwidth}{!}{
\begin{tabular}{|c|c|}
\hline
\makecell{\textbf{Performance Evaluation}\\ \textbf{Parameter}}& \makecell{\textbf{Mathematical }\\ \textbf{Equation}}\\
\hline
Accuracy & $\frac{TP + TN}{FP + FN + TP + TN}$ \\
\hline
Precision & $\frac{TP}{FP + TP}$ \\
\hline
Recall & $\frac{TP}{FN + TP}$ \\
\hline
F1-Score & $2\frac{Prec \times Sens}{Prec + Sens}$ \\
\hline
AUC & \makecell{Area Under\\ROC Curve}\\
\hline

\end{tabular}}
\end{table}

\subsection{Performaces}
\begin{figure}[!h]
\centering

\begin{subfigure}[]
    \centering
    \includegraphics[width=0.8\textwidth]{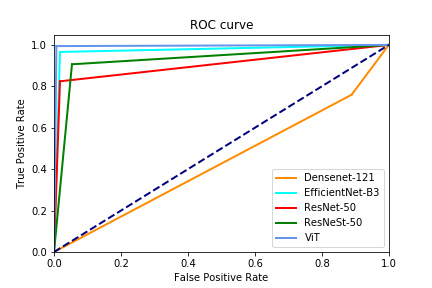}
\end{subfigure}

\begin{subfigure}[]
    \centering
    \includegraphics[width=0.8\textwidth]{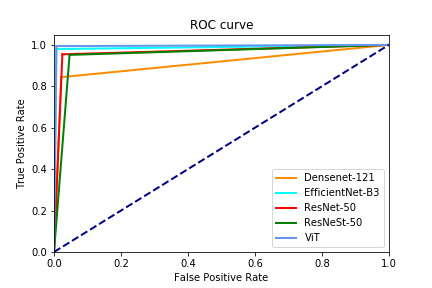}
\end{subfigure}

\caption{ROC curve of networks without CycleGAN (a) and with it (b).\label{figroc}}
\end{figure}

This part of the paper is dedicated to showing the results of networks. Each network is first trained without using the CycleGAN, and then the effect of adding CycleGAN is measured. Tables \ref{tabnogan} and \ref{tabgan} demonstrate the network results without and with CycleGAN, and Figures \ref{fignnogan} and \ref{figngan} also show the networks' learning curves. To make the results reliable, each network is evaluated ten times, and then the mean of performances, with confidence intervals, are presented. As observable in these tables, CycleGAN has improved the performance of EfficientNet, Resnet, and ResNeSt dramatically. Nevertheless, the ViT results show no sign of improvement in the presence of CycleGAN; this is arguably due to its robustness or indistinguishability of wrongly classified samples from the other class. ROC curve for one run of the networks is also plotted in Figure \ref{figroc}.

\begin{table}[h] 
\caption{Results without CycleGAN.\label{tabnogan}}
\centering
\resizebox{\textwidth}{!}{
\begin{tabular}{|c|c|c|c|c|c|}
\hline
\textbf{Network} & \textbf{Accuracy (\%)}	& \textbf{Precision (\%)}	&\textbf{Recall (\%)}	&\textbf{F1-score (\%)} & \textbf{AUC (\%)}\\
\hline
Densenet-121    & 88.05 ± 3.81 & 80.94 ± 5.55 & 93.18 ± 3.70 & 87.36 ± 3.63 & 96.71 ± 2.80 \\ \hline
EfficientNet-B3 & 94.69 ± 2.04 & 92.55 ± 3.40 & 96.77 ± 1.57 & 94.09 ± 2.19 & 99.03 ± 0.66 \\ \hline
Resnet-50       & 94.69 ± 1.15 & 90.31 ± 2.21 & 98.74 ± 0.48 & 94.25 ± 1.15 & 99.43 ± 0.20 \\ \hline
ResNeSt-50      & 96.30 ± 2.31 & 93.96 ± 3.00 & 98.02 ± 2.43 & 95.97 ± 2.53 & 99.60 ± 1.18 \\ \hline
ViT             & 99.60 ± 0.79 & 99.46 ± 1.39 & 99.64 ± 0.38 & 99.55 ± 0.88 & 99.99 ± 0.10 \\ \hline
\end{tabular}}
\end{table}

\begin{table}[h] 
\caption{Results with CycleGAN.\label{tabgan}}
\centering
\resizebox{\textwidth}{!}{
\begin{tabular}{|c|c|c|c|c|c|}
\hline
\textbf{Network} & \textbf{Accuracy (\%)}	& \textbf{Precision (\%)}	&\textbf{Recall (\%)}	&\textbf{F1-score (\%)} & \textbf{AUC (\%)}\\
\hline
Densenet-121    & 89.24 ± 3.78 & 81.67 ± 5.84 & 96.23 ± 2.45 & 88.64 ± 3.55 & 97.22 ± 1.89 \\ \hline
EfficientNet-B3 & 98.25 ± 2.57 & 97.03 ± 3.51 & 99.28 ± 2.20 & 98.05 ± 2.80 & 99.79 ± 0.90 \\ \hline
Resnet-50       & 96.20 ± 0.79 & 94.09 ± 1.33 & 97.49 ± 1.11 & 95.78 ± 0.87 & 99.43 ± 0.42 \\ \hline
ResNeSt-50      & 98.89 ± 1.09 & 98.58 ± 1.34 & 99.10 ± 1.70 & 98.75 ± 1.24 & 99.95 ± 0.22 \\ \hline
ViT             & 99.20 ± 2.91 & 98.92 ± 3.97 & 98.92 ± 2.41 & 99.10 ± 3.19 & 99.95 ± 0.92 \\ \hline

\end{tabular}}
\end{table}
\clearpage
\begin{landscape}
\setlength\LTleft{-108pt}            
\setlength\LTright{0pt}           
\pagestyle{empty}

\begin{figure}[h]
\centering
\begin{subfigure}[]
    \centering
    \includegraphics[width=0.6\textwidth]{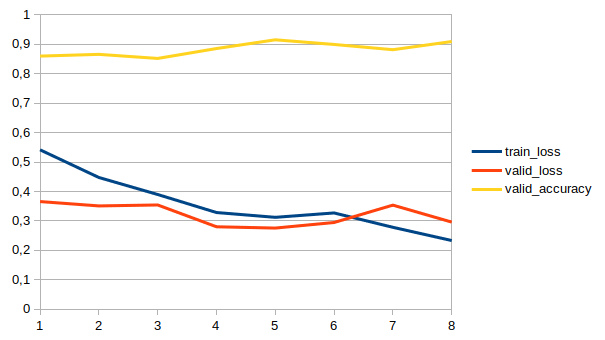}
\end{subfigure}
\hspace{2pt}
\begin{subfigure}[]
    \centering
    \includegraphics[width=0.6\textwidth]{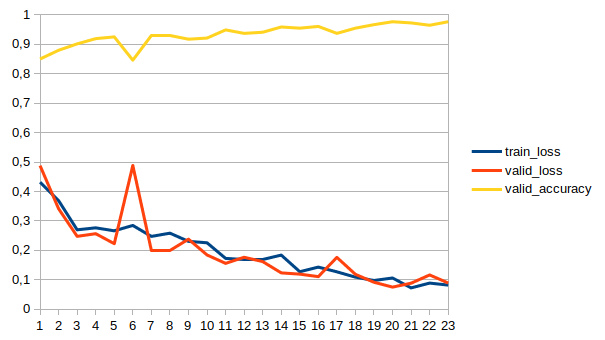}
\end{subfigure}

\begin{subfigure}[]
    \centering
    \includegraphics[width=0.6\textwidth]{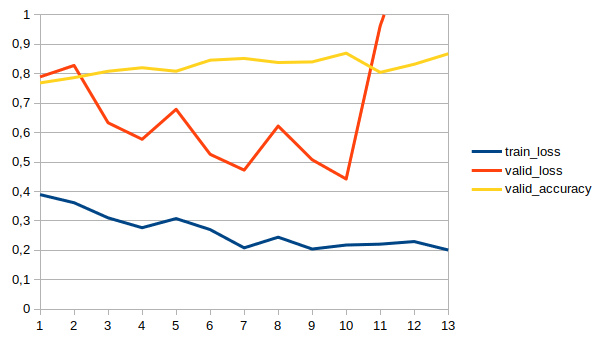}
\end{subfigure}
\hspace{2pt}
\begin{subfigure}[]
    \centering
    \includegraphics[width=0.6\textwidth]{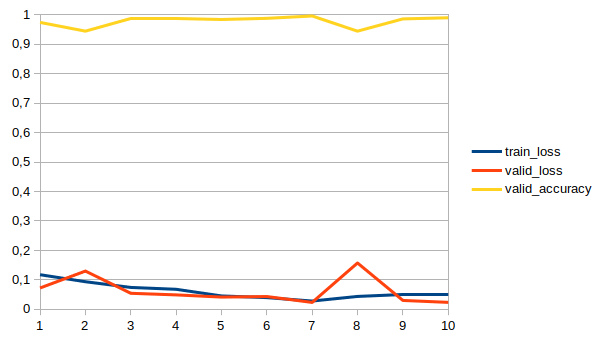}
\end{subfigure}

\begin{subfigure}[]
    \centering
    \includegraphics[width=0.6\textwidth]{pics/resnest.jpg}
\end{subfigure}

\caption{Learning curve of networks without CycleGAN for (a) DenseNet, (b) EfficientNet, (c) ResNet, (d) ViT, and (e) ResNeSt.\label{fignnogan}}

\end{figure}

\begin{figure}[!h]
\centering
\begin{subfigure}[]
    \centering
    \includegraphics[width=0.6\textwidth]{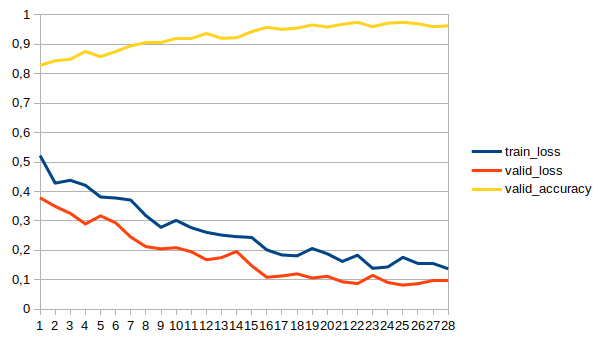}
\end{subfigure}
\hspace{2pt}
\begin{subfigure}[]
    \centering
    \includegraphics[width=0.6\textwidth]{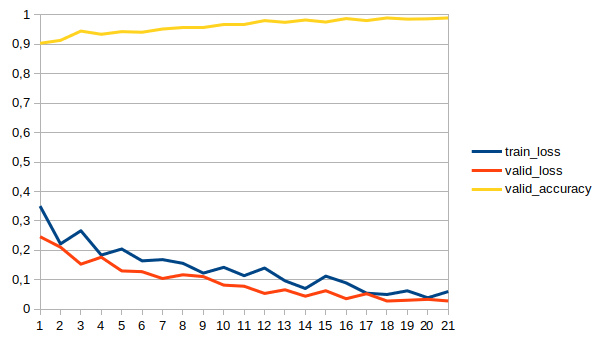}
\end{subfigure}

\begin{subfigure}[]
    \centering
    \includegraphics[width=0.6\textwidth]{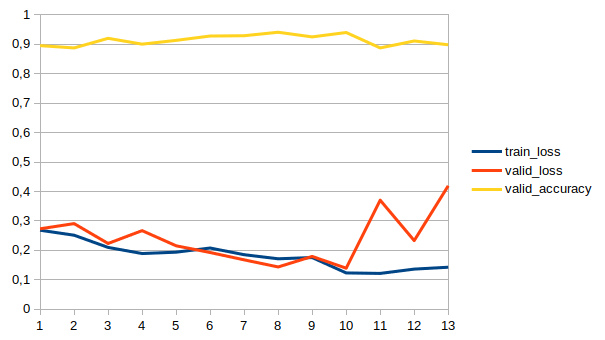}
\end{subfigure}
\hspace{2pt}
\begin{subfigure}[]
    \centering
    \includegraphics[width=0.6\textwidth]{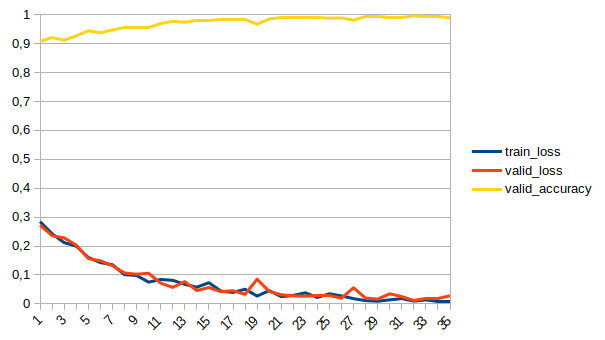}
\end{subfigure}

\begin{subfigure}[]
    \centering
    \includegraphics[width=0.6\textwidth]{pics/resnest-cyclegan.jpg}
\end{subfigure}

\caption{Learning curve of networks with CycleGAN for (a) DenseNet, (b) EfficientNet, (c) ResNet, (d) ViT, and (e) ResNeSt.\label{figngan}}
\end{figure}
\end{landscape}

\section{Discussion}

In recent years, convolutional neural networks have revolutionized the field of image processing. Medical diagnoses are no exception, and today in numerous research papers in this field, the use of these networks to achieve the best accuracy is seen. Diagnosis of COVID-19 disease from CT images is also one of the applications of these networks. In this article, the performance of different networks in this task was examined, and also by applying a new method, an attempt was made to improve the performance of these networks. The networks used in this paper were Resnet, EfficientNet, Densenet, ViT, and ResNest, and the data augmentation method was based on CycleGAN. Table \ref{tab:newt} summarizes the proposed method of previous papers. By comparing this table with our current work, the advantages of our work can be listed as using ViT, a transformer-based architecture that has achieved state-of-the-art performances; collecting a new dataset; and finally using CycleGAN for data augmentation.

\begin{table}[h]
    \caption{Summary of related works.}
    \label{tab:newt}
    \centering
    
    \resizebox{\textwidth}{!}{
    \begin{tabular}{|c|c|c|c|c|c|}
        \hline
\textbf{Ref} & \textbf{Dataset} & \textbf{Number of Cases (Images)} & \textbf{Pre-Processing} & \textbf{DNN}  & \textbf{Performance (\%)}\\ \hline
\cite{a1} & SIRM & 3000 COVID-19, 3000 HC & Patches Extraction & PreTrain Networks & Acc=98.27 \\ \hline
\cite{a2} & Zhao et al & 460 COVID-19, 397 HC   & DA & SqueezeNet & Acc=85.03 \\ \hline
\cite{a3} & Indian & 2373 COVID-19, 6321 HC & -- & Ensemble DCCNs & Acc=98.83 \\ \hline
\cite{a4} & Different Datasets & -- & Visual Inspection & BigBiGAN & -- \\ \hline
\cite{a5} & Clinical & 148 COVID-19, 148 HC   & Visual Inspection & ResGNet-C & Acc=96.62 \\ \hline
\cite{a6} & Clinical & 349 COVID-19, 397 HC   & Scaling Process, DA & MKs-ELM-DNN & Acc=98.36 \\ \hline
\cite{a7} & COVID-CT & -- & DA & U-net + DCN & Acc=92.87 \\ \hline
\cite{a8} & Public Dataset & 2933 COVID-19 & Normalization, Resizing & EDL\_COVID & Acc= 99.054 \\ \hline
\cite{a9} & Clinical & 320 COVID-19, 320 HC   & HS, Margin Crop, Resizing & FGCNet & Acc=97.14   \\ \hline
\cite{a10} & Clinical & -- & ROIs Extraction & Modified Inception & Acc=89.5 \\ \hline
\cite{a11} & Clinical & 3389 COVID-19, 1593 HC & Standard Preprocessing & 3D ResNet34 & Acc=87.5 \\ \hline
\cite{a12} & COVIDx-CT & 104,009 & DA & COVIDNet-CT & Acc= 99.1 \\ \hline
\cite{a13} & Different Datasets & 349 COVID-19, 397 HC   & Resizing, Normalization, DA & ResNet18 & Acc=99.4    \\ \hline
\cite{a14} & COVID-CT & 345 COVID-19, 397 HC   & Resizing, DA & CGAN + ResNet50 & Acc=82.91   \\ \hline
\cite{a15} & Clinical & -- & Resizing, Padding, DA & 3D-CNN & AUC=70 \\ \hline
\cite{a16} & SARS-CoV-2 & 1252 COVID-19, 1230 HC & -- & GAN with WOA + InceptionV3 & Acc=99.22   \\ \hline
\cite{a17} & Different Datasets & -- & Resizing & Inception V1 & Acc=95.78 \\ \hline
\cite{a18} & Clinical & 2267 COVID-19, 1235 HC & Normalization, Cropping, Resizing & ResNet50 & Acc=93 \\ \hline
\cite{a19} & Clinical & -- & \makecell{Visual Inspection, ROI, Cropping\\and Resizing} & ResNet 101 & Acc=99.51   \\ \hline
\cite{a20} & Different Datasets & 413 COVID-19, 439 HC   & -- & ResNet-50 + 3D-CNN & Acc=93.01 \\ \hline
\cite{a21} & Clinical & -- & DA & Multi-View U-Net + 3D-CNN & Acc=90.6 \\ \hline
\cite{a22} & Different Datasets & -- & Resized, Intensity Normalized & FCN & Acc=94.67 \\ \hline
\cite{a23} & Zhao et al & 746 & -- & GAN + ShuffleNet & Acc=84.9 \\ \hline
\cite{a24} & COVID-CT & 345 COVID-19,  375 HC  & 2D RDWT, Resizing & ResNet50 & Acc=92.2 \\ \hline
\cite{a25} & SARS-CoV-2 & 1262 COVID-19, 1230 HC & -- & CSVM & Acc=94.03 \\ \hline
\cite{a26} & Different Datasets & 767 & -- & Different PreTrain Methods & Acc=75 \\ \hline
\cite{a27} & MedSeg DII & -- & Resizing & U-Net + VGG16 and Resnet-50 & Acc=99.4 \\ \hline
\cite{a28} & Basrah & 1425 & Resizing & VGG 16 & Acc=99 \\ \hline
\cite{a29} & Kaggle & 1252 COVID, 1240 HC    & Resizing, Normalization, DA & Covid CT-net & Acc=95.78 \\ \hline
\cite{a30} & COVID-CT & 312 COVID-19, 396 HC   & Normalization & LeNet-5 & Acc=95.07 \\ \hline
Ours & Clinical & 1766 COVID-19, 1397 HC & Filtering, DA using CycleGAN & Different PreTrain Methods & Acc=99.60\\ \hline
\end{tabular}}  
\end{table}

Eventually, the ViT network reached an accuracy of 99.60\%, which shows its state-of-the-art performance and proves that it can be used as the heart of a CADS. By comparing the performances of our method compared to previous works in table \ref{tab1}, our methods' superiority is quite observable. The advantages of adding CycleGAN were also clearly displayed, and it was shown that this method could be used for this task by data augmentation to improve the performance of most deep neural networks. Finally, this article's achievements can be summarized in: first, introducing a new database and its public release, second, examining the performance of various neural networks on this database, and finally evaluating the use of CycleGAN for data augmentation and its impact on networks performances. Additionally, the performance of ViT was never previously studied for this task, which was investigated in this paper as well. To evaluate the method, a CT scan dataset was collected by physicians, which we also made available to researchers in public. Also, due to the fact that this dataset was collected from people suspected of having COVID-19, normal class data, unlike many previous datasets in this field, were collected from patients with suspicious symptoms and not from other diseases.

\section{Conclusion, and Future Works}

In the past year or so, nearly all people have found their lives changed due to the COVID-19 outbreak. Researchers in image processing and machine learning have not been an exception, considering many research papers that have been published on a variety of automatic diagnostic methods using medical imaging modalities and machine learning methods. Building an accurate diagnostic system in these pandemic conditions can relieve many of the burdens on physicians and also help to improve the situation. In this paper, the use of convolutional neural networks for the task at hand was investigated, and also the effect of adding CycleGAN for data augmentation was examined as another novelty of the paper. Finally, our method reached state-of-the-art performances and also have outperformed prior works, which shows its superiority.

For future work, several different paths can be considered; first, more complicated methods in deep neural networks can be used, such as deep metric, few-shot learning, or feature fusion solutions. Also, the combination of different datasets to improve the accuracy and evaluate its impact on the training of different networks can be examined. Finally, combining different modalities to increase accuracy can also be a direction for future research.

\bibliography{main}

\begin{thebibliography}{10}
\expandafter\ifx\csname url\endcsname\relax
  \def\url#1{\texttt{#1}}\fi
\expandafter\ifx\csname urlprefix\endcsname\relax\def\urlprefix{URL }\fi
\expandafter\ifx\csname href\endcsname\relax
  \def\href#1#2{#2} \def\path#1{#1}\fi

\bibitem{i1}
M.~Jamshidi, A.~Lalbakhsh, J.~Talla, Z.~Peroutka, F.~Hadjilooei, P.~Lalbakhsh,
  M.~Jamshidi, L.~La~Spada, M.~Mirmozafari, M.~Dehghani, et~al., Artificial
  intelligence and covid-19: deep learning approaches for diagnosis and
  treatment, IEEE Access 8 (2020) 109581--109595.
\newblock \href {http://dx.doi.org/10.1109/ACCESS.2020.3001973}
  {\path{doi:10.1109/ACCESS.2020.3001973}}.

\bibitem{i2}
S.~Vaid, R.~Kalantar, M.~Bhandari, Deep learning covid-19 detection bias:
  accuracy through artificial intelligence, International Orthopaedics 44
  (2020) 1539--1542.
\newblock \href {http://dx.doi.org/10.1007/s00264-020-04609-7}
  {\path{doi:10.1007/s00264-020-04609-7}}.

\bibitem{i3i4}
G.~A. Perchetti, A.~K. Nalla, M.-L. Huang, H.~Zhu, Y.~Wei, L.~Stensland, M.~A.
  Loprieno, K.~R. Jerome, A.~L. Greninger, Validation of sars-cov-2 detection
  across multiple specimen types, Journal of clinical virology 128 (2020)
  104438.
\newblock \href {http://dx.doi.org/10.1016/j.jcv.2020.104438}
  {\path{doi:10.1016/j.jcv.2020.104438}}.

\bibitem{i5}
A.~Lopez-Rincon, A.~Tonda, L.~Mendoza-Maldonado, D.~G. Mulders, R.~Molenkamp,
  C.~A. Perez-Romero, E.~Claassen, J.~Garssen, A.~D. Kraneveld, Classification
  and specific primer design for accurate detection of sars-cov-2 using deep
  learning, Scientific reports 11~(1) (2021) 1--11.
\newblock \href {http://dx.doi.org/10.1038/s41598-020-80363-5}
  {\path{doi:10.1038/s41598-020-80363-5}}.

\bibitem{i6}
A.~Shoeibi, M.~Khodatars, R.~Alizadehsani, N.~Ghassemi, M.~Jafari, P.~Moridian,
  A.~Khadem, D.~Sadeghi, S.~Hussain, A.~Zare, et~al., Automated detection and
  forecasting of covid-19 using deep learning techniques: A review, arXiv
  preprint arXiv:2007.10785.

\bibitem{i7}
A.~Kumar, P.~K. Gupta, A.~Srivastava, A review of modern technologies for
  tackling covid-19 pandemic, Diabetes \& Metabolic Syndrome: Clinical Research
  \& Reviews 14~(4) (2020) 569--573.
\newblock \href {http://dx.doi.org/10.1016/j.dsx.2020.05.008}
  {\path{doi:10.1016/j.dsx.2020.05.008}}.

\bibitem{nnone}
M.~Mohammadpoor, A.~Shoeibi, H.~Shojaee, et~al., A hierarchical classification
  method for breast tumor detection, Iranian Journal of Medical Physics 13~(4)
  (2016) 261--268.
\newblock \href {http://dx.doi.org/10.22038/IJMP.2016.8453}
  {\path{doi:10.22038/IJMP.2016.8453}}.

\bibitem{i8}
O.~Albahri, A.~Zaidan, A.~Albahri, B.~Zaidan, K.~H. Abdulkareem, Z.~Al-Qaysi,
  A.~Alamoodi, A.~Aleesa, M.~Chyad, R.~Alesa, et~al., Systematic review of
  artificial intelligence techniques in the detection and classification of
  covid-19 medical images in terms of evaluation and benchmarking: Taxonomy
  analysis, challenges, future solutions and methodological aspects, Journal of
  infection and public health\href
  {http://dx.doi.org/10.1016/j.jiph.2020.06.028}
  {\path{doi:10.1016/j.jiph.2020.06.028}}.

\bibitem{i9}
A.~Assiri, A.~McGeer, T.~M. Perl, C.~S. Price, A.~A. Al~Rabeeah, D.~A.
  Cummings, Z.~N. Alabdullatif, M.~Assad, A.~Almulhim, H.~Makhdoom, et~al.,
  Hospital outbreak of middle east respiratory syndrome coronavirus, New
  England Journal of Medicine 369~(5) (2013) 407--416.
\newblock \href {http://dx.doi.org/10.1056/NEJMoa1306742}
  {\path{doi:10.1056/NEJMoa1306742}}.

\bibitem{i10}
S.~Khan, B.~Shaker, S.~Ahmad, S.~W. Abbasi, M.~Arshad, A.~Haleem, S.~Ismail,
  A.~Zaib, W.~Sajjad, Towards a novel peptide vaccine for middle east
  respiratory syndrome coronavirus and its possible use against pandemic
  covid-19, Journal of molecular liquids 324 (2021) 114706.
\newblock \href {http://dx.doi.org/10.1016/j.molliq.2020.114706}
  {\path{doi:10.1016/j.molliq.2020.114706}}.

\bibitem{i11}
S.~Lalmuanawma, J.~Hussain, L.~Chhakchhuak, Applications of machine learning
  and artificial intelligence for covid-19 (sars-cov-2) pandemic: A review,
  Chaos, Solitons \& Fractals (2020) 110059\href
  {http://dx.doi.org/10.1016/j.chaos.2020.110059}
  {\path{doi:10.1016/j.chaos.2020.110059}}.

\bibitem{i12}
D.~J. Alouani, R.~R. Rajapaksha, M.~Jani, D.~D. Rhoads, N.~Sadri, Deep learning
  analysis improves specificity of sars-cov-2 real time pcr, Journal of
  Clinical Microbiology\href {http://dx.doi.org/10.1128/JCM.02959-20}
  {\path{doi:10.1128/JCM.02959-20}}.

\bibitem{i13}
N.~Benameur, R.~Mahmoudi, S.~Zaid, Y.~Arous, B.~Hmida, M.~H. Bedoui, Sars-cov-2
  diagnosis using medical imaging techniques and artificial intelligence: A
  review, Clinical Imaging\href
  {http://dx.doi.org/10.1016/j.clinimag.2021.01.019}
  {\path{doi:10.1016/j.clinimag.2021.01.019}}.

\bibitem{i14}
B.~Ghoshal, A.~Tucker, Estimating uncertainty and interpretability in deep
  learning for coronavirus (covid-19) detection, arXiv preprint
  arXiv:2003.10769.

\bibitem{i15}
D.~Sharifrazi, R.~Alizadehsani, M.~Roshanzamir, J.~H. Joloudari, A.~Shoeibi,
  M.~Jafari, S.~Hussain, Z.~A. Sani, F.~Hasanzadeh, F.~Khozeimeh, et~al.,
  Fusion of convolution neural network, support vector machine and sobel filter
  for accurate detection of covid-19 patients using x-ray images, Biomedical
  Signal Processing and Control (2021) 102622\href
  {http://dx.doi.org/10.1016/j.bspc.2021.102622}
  {\path{doi:10.1016/j.bspc.2021.102622}}.

\bibitem{i16}
H.~B. Syeda, M.~Syed, K.~W. Sexton, S.~Syed, S.~Begum, F.~Syed, F.~Prior,
  F.~Yu~Jr, Role of machine learning techniques to tackle the covid-19 crisis:
  Systematic review, JMIR medical informatics 9~(1) (2021) e23811.
\newblock \href {http://dx.doi.org/10.2196/23811} {\path{doi:10.2196/23811}}.

\bibitem{i17}
D.~Dong, Z.~Tang, S.~Wang, H.~Hui, L.~Gong, Y.~Lu, Z.~Xue, H.~Liao, F.~Chen,
  F.~Yang, et~al., The role of imaging in the detection and management of
  covid-19: a review, IEEE reviews in biomedical engineering\href
  {http://dx.doi.org/10.1109/RBME.2020.2990959}
  {\path{doi:10.1109/RBME.2020.2990959}}.

\bibitem{i18}
A.~Albahri, R.~A. Hamid, J.~K. Alwan, Z.~Al-Qays, A.~Zaidan, B.~Zaidan,
  A.~Albahri, A.~AlAmoodi, J.~M. Khlaf, E.~Almahdi, et~al., Role of biological
  data mining and machine learning techniques in detecting and diagnosing the
  novel coronavirus (covid-19): a systematic review, Journal of medical systems
  44 (2020) 1--11.
\newblock \href {http://dx.doi.org/10.1007/s10916-020-01582-x}
  {\path{doi:10.1007/s10916-020-01582-x}}.

\bibitem{i19}
A.~Tahamtan, A.~Ardebili, Real-time rt-pcr in covid-19 detection: issues
  affecting the results, Expert review of molecular diagnostics 20~(5) (2020)
  453--454.
\newblock \href {http://dx.doi.org/10.1080/14737159.2020.1757437}
  {\path{doi:10.1080/14737159.2020.1757437}}.

\bibitem{i20}
L.~Lan, D.~Xu, G.~Ye, C.~Xia, S.~Wang, Y.~Li, H.~Xu, Positive rt-pcr test
  results in patients recovered from covid-19, Jama 323~(15) (2020) 1502--1503.
\newblock \href {http://dx.doi.org/10.1001/jama.2020.2783}
  {\path{doi:10.1001/jama.2020.2783}}.

\bibitem{i21}
Y.~Fang, H.~Zhang, J.~Xie, M.~Lin, L.~Ying, P.~Pang, W.~Ji, Sensitivity of
  chest ct for covid-19: comparison to rt-pcr, Radiology 296~(2) (2020)
  E115--E117.
\newblock \href {http://dx.doi.org/10.1148/radiol.2020200432}
  {\path{doi:10.1148/radiol.2020200432}}.

\bibitem{i22}
T.~Ozturk, M.~Talo, E.~A. Yildirim, U.~B. Baloglu, O.~Yildirim, U.~R. Acharya,
  Automated detection of covid-19 cases using deep neural networks with x-ray
  images, Computers in biology and medicine 121 (2020) 103792.
\newblock \href {http://dx.doi.org/10.1016/j.compbiomed.2020.103792}
  {\path{doi:10.1016/j.compbiomed.2020.103792}}.

\bibitem{i23}
V.~Shah, R.~Keniya, A.~Shridharani, M.~Punjabi, J.~Shah, N.~Mehendale,
  Diagnosis of covid-19 using ct scan images and deep learning techniques,
  Emergency radiology (2021) 1--9\href
  {http://dx.doi.org/10.1007/s10140-020-01886-y}
  {\path{doi:10.1007/s10140-020-01886-y}}.

\bibitem{i24}
S.~R. Nayak, D.~R. Nayak, U.~Sinha, V.~Arora, R.~B. Pachori, Application of
  deep learning techniques for detection of covid-19 cases using chest x-ray
  images: A comprehensive study, Biomedical Signal Processing and Control 64
  (2021) 102365.
\newblock \href {http://dx.doi.org/10.1016/j.bspc.2020.102365}
  {\path{doi:10.1016/j.bspc.2020.102365}}.

\bibitem{i25}
S.~Karakanis, G.~Leontidis, Lightweight deep learning models for detecting
  covid-19 from chest x-ray images, Computers in Biology and Medicine 130
  (2021) 104181.
\newblock \href {http://dx.doi.org/10.1016/j.compbiomed.2020.104181}
  {\path{doi:10.1016/j.compbiomed.2020.104181}}.

\bibitem{i26}
M.~Blain, M.~T. Kassin, N.~Varble, X.~Wang, Z.~Xu, D.~Xu, G.~Carrafiello,
  V.~Vespro, E.~Stellato, A.~M. Ierardi, et~al., Determination of disease
  severity in covid-19 patients using deep learning in chest x-ray images,
  Diagnostic and Interventional Radiology 27~(1) (2021) 20.
\newblock \href {http://dx.doi.org/10.5152/dir.2020.20205}
  {\path{doi:10.5152/dir.2020.20205}}.

\bibitem{i27}
E.~Hussain, M.~Hasan, M.~A. Rahman, I.~Lee, T.~Tamanna, M.~Z. Parvez, Corodet:
  A deep learning based classification for covid-19 detection using chest x-ray
  images, Chaos, Solitons \& Fractals 142 (2021) 110495.
\newblock \href {http://dx.doi.org/10.1016/j.chaos.2020.110495}
  {\path{doi:10.1016/j.chaos.2020.110495}}.

\bibitem{i28}
N.~Lassau, S.~Ammari, E.~Chouzenoux, H.~Gortais, P.~Herent, M.~Devilder,
  S.~Soliman, O.~Meyrignac, M.-P. Talabard, J.-P. Lamarque, et~al., Integrating
  deep learning ct-scan model, biological and clinical variables to predict
  severity of covid-19 patients, Nature communications 12~(1) (2021) 1--11.
\newblock \href {http://dx.doi.org/10.1038/s41467-020-20657-4}
  {\path{doi:10.1038/s41467-020-20657-4}}.

\bibitem{i29}
R.~Alizadehsani, D.~Sharifrazi, N.~H. Izadi, J.~H. Joloudari, A.~Shoeibi, J.~M.
  Gorriz, S.~Hussain, J.~E. Arco, Z.~A. Sani, F.~Khozeimeh, et~al.,
  Uncertainty-aware semi-supervised method using large unlabelled and limited
  labeled covid-19 data, arXiv preprint arXiv:2102.06388.

\bibitem{i30}
S.~Wang, B.~Kang, J.~Ma, X.~Zeng, M.~Xiao, J.~Guo, M.~Cai, J.~Yang, Y.~Li,
  X.~Meng, et~al., A deep learning algorithm using ct images to screen for
  corona virus disease (covid-19), European Radiology (2021) 1--9\href
  {http://dx.doi.org/10.1007/s00330-021-07715-1}
  {\path{doi:10.1007/s00330-021-07715-1}}.

\bibitem{i31}
S.~Hu, Y.~Gao, Z.~Niu, Y.~Jiang, L.~Li, X.~Xiao, M.~Wang, E.~F. Fang,
  W.~Menpes-Smith, J.~Xia, et~al., Weakly supervised deep learning for covid-19
  infection detection and classification from ct images, IEEE Access 8 (2020)
  118869--118883.
\newblock \href {http://dx.doi.org/10.1109/ACCESS.2020.3005510}
  {\path{doi:10.1109/ACCESS.2020.3005510}}.

\bibitem{i32}
M.~M. Islam, F.~Karray, R.~Alhajj, J.~Zeng, A review on deep learning
  techniques for the diagnosis of novel coronavirus (covid-19), IEEE Access 9
  (2021) 30551--30572.
\newblock \href {http://dx.doi.org/10.1109/ACCESS.2021.3058537}
  {\path{doi:10.1109/ACCESS.2021.3058537}}.

\bibitem{i33}
A.~W. Salehi, P.~Baglat, G.~Gupta, Review on machine and deep learning models
  for the detection and prediction of coronavirus, Materials Today: Proceedings
  33 (2020) 3896--3901.
\newblock \href {http://dx.doi.org/10.1016/j.matpr.2020.06.245}
  {\path{doi:10.1016/j.matpr.2020.06.245}}.

\bibitem{abc4}
P.~K. Chaudhary, R.~B. Pachori, Automatic diagnosis of covid-19 and pneumonia
  using fbd method, in: 2020 IEEE International Conference on Bioinformatics
  and Biomedicine (BIBM), IEEE, 2020, pp. 2257--2263.
\newblock \href {http://dx.doi.org/10.1109/BIBM49941.2020.9313252}
  {\path{doi:10.1109/BIBM49941.2020.9313252}}.

\bibitem{i34}
H.~Swapnarekha, H.~S. Behera, J.~Nayak, B.~Naik, Role of intelligent computing
  in covid-19 prognosis: A state-of-the-art review, Chaos, Solitons \& Fractals
  138 (2020) 109947.
\newblock \href {http://dx.doi.org/10.1016/j.chaos.2020.109947}
  {\path{doi:10.1016/j.chaos.2020.109947}}.

\bibitem{i35}
F.~Shi, J.~Wang, J.~Shi, Z.~Wu, Q.~Wang, Z.~Tang, K.~He, Y.~Shi, D.~Shen,
  Review of artificial intelligence techniques in imaging data acquisition,
  segmentation and diagnosis for covid-19, IEEE reviews in biomedical
  engineering\href {http://dx.doi.org/10.1109/RBME.2020.2987975}
  {\path{doi:10.1109/RBME.2020.2987975}}.

\bibitem{r10}
J.-Y. Zhu, T.~Park, P.~Isola, A.~A. Efros, Unpaired image-to-image translation
  using cycle-consistent adversarial networks, in: Proceedings of the IEEE
  international conference on computer vision, 2017, pp. 2223--2232.

\bibitem{abc5}
A.~Bar-El, D.~Cohen, N.~Cahan, H.~Greenspan, Improved cyclegan with application
  to covid-19 classification, in: Medical Imaging 2021: Image Processing, Vol.
  11596, International Society for Optics and Photonics, 2021, p. 1159614.
\newblock \href {http://dx.doi.org/10.1117/12.2582162}
  {\path{doi:10.1117/12.2582162}}.

\bibitem{abc3}
N.~Ghassemi, H.~Mahami, M.~T. Darbandi, A.~Shoeibi, S.~Hussain, F.~Nasirzadeh,
  R.~Alizadehsani, D.~Nahavandi, A.~Khosravi, S.~Nahavandi, Material
  recognition for automated progress monitoring using deep learning methods,
  arXiv preprint arXiv:2006.16344.

\bibitem{r4}
G.~Huang, Z.~Liu, L.~Van Der~Maaten, K.~Q. Weinberger, Densely connected
  convolutional networks, in: Proceedings of the IEEE conference on computer
  vision and pattern recognition, 2017, pp. 4700--4708.

\bibitem{r2}
K.~He, X.~Zhang, S.~Ren, J.~Sun, Deep residual learning for image recognition,
  in: Proceedings of the IEEE conference on computer vision and pattern
  recognition, 2016, pp. 770--778.
\newblock \href {http://dx.doi.org/10.1109/CVPR.2016.90}
  {\path{doi:10.1109/CVPR.2016.90}}.

\bibitem{r5}
H.~Zhang, C.~Wu, Z.~Zhang, Y.~Zhu, Z.~Zhang, H.~Lin, Y.~Sun, T.~He, J.~Mueller,
  R.~Manmatha, et~al., Resnest: Split-attention networks, arXiv preprint
  arXiv:2004.08955.

\bibitem{r6}
A.~Dosovitskiy, L.~Beyer, A.~Kolesnikov, D.~Weissenborn, X.~Zhai,
  T.~Unterthiner, M.~Dehghani, M.~Minderer, G.~Heigold, S.~Gelly, et~al., An
  image is worth 16x16 words: Transformers for image recognition at scale,
  arXiv preprint arXiv:2010.11929.

\bibitem{ndeep}
I.~Goodfellow, Y.~Bengio, A.~Courville, Y.~Bengio, Deep learning, Vol.~1, MIT
  press Cambridge, 2016.

\bibitem{abc2}
J.~M. G{\'o}rriz, J.~Ram{\'\i}rez, A.~Ort{\'\i}z, F.~J. Mart{\'\i}nez-Murcia,
  F.~Segovia, J.~Suckling, M.~Leming, Y.-D. Zhang, J.~R.
  {\'A}lvarez-S{\'a}nchez, G.~Bologna, et~al., Artificial intelligence within
  the interplay between natural and artificial computation: Advances in data
  science, trends and applications, Neurocomputing 410 (2020) 237--270.
\newblock \href {http://dx.doi.org/10.1016/j.neucom.2020.05.078}
  {\path{doi:10.1016/j.neucom.2020.05.078}}.

\bibitem{nbishop}
C.~M. Bishop, Pattern recognition and machine learning, springer, 2006.

\bibitem{nntwo}
R.~Alizadehsani, Z.~Alizadeh~Sani, M.~Behjati, Z.~Roshanzamir, S.~Hussain,
  N.~Abedini, F.~Hasanzadeh, A.~Khosravi, A.~Shoeibi, M.~Roshanzamir, et~al.,
  Risk factors prediction, clinical outcomes, and mortality in covid-19
  patients, Journal of medical virology 93~(4) (2021) 2307--2320.
\newblock \href {http://dx.doi.org/10.1002/jmv.26699}
  {\path{doi:10.1002/jmv.26699}}.

\bibitem{abc1}
C.~Jim{\'e}nez-Mesa, J.~Ram{\'\i}rez, J.~Suckling, J.~V{\"o}glein, J.~Levin,
  J.~M. G{\'o}rriz, A.~D. N.~I. ADNI, D.~I. A.~N. DIAN, Deep learning in
  current neuroimaging: a multivariate approach with power and type i error
  control but arguable generalization ability, arXiv preprint arXiv:2103.16685.

\bibitem{a1}
U.~{\"O}zkaya, {\c{S}}.~{\"O}zt{\"u}rk, M.~Barstugan, Coronavirus (covid-19)
  classification using deep features fusion and ranking technique, in: Big Data
  Analytics and Artificial Intelligence Against COVID-19: Innovation Vision and
  Approach, Springer, 2020, pp. 281--295.
\newblock \href {http://dx.doi.org/10.1007/978-3-030-55258-9\_17}
  {\path{doi:10.1007/978-3-030-55258-9\_17}}.

\bibitem{a2}
M.~Polsinelli, L.~Cinque, G.~Placidi, A light cnn for detecting covid-19 from
  ct scans of the chest, Pattern Recognition Letters 140 (2020) 95--100.
\newblock \href {http://dx.doi.org/10.1016/j.patrec.2020.10.001}
  {\path{doi:10.1016/j.patrec.2020.10.001}}.

\bibitem{a17}
K.~Yang, X.~Liu, Y.~Yang, X.~Liao, R.~Wang, X.~Zeng, Y.~Wang, M.~Zhang,
  T.~Zhang, End-to-end covid-19 screening with 3d deep learning on chest
  computed tomography.

\bibitem{a8}
T.~Zhou, H.~Lu, Z.~Yang, S.~Qiu, B.~Huo, Y.~Dong, The ensemble deep learning
  model for novel covid-19 on ct images, Applied Soft Computing 98 (2021)
  106885.
\newblock \href {http://dx.doi.org/10.1016/j.asoc.2020.106885}
  {\path{doi:10.1016/j.asoc.2020.106885}}.

\bibitem{a3}
D.~Singh, V.~Kumar, M.~Kaur, Densely connected convolutional networks-based
  covid-19 screening model, Applied Intelligence (2021) 1--8\href
  {http://dx.doi.org/10.1007/s10489-020-02149-6}
  {\path{doi:10.1007/s10489-020-02149-6}}.

\bibitem{a4}
J.~Song, H.~Wang, Y.~Liu, W.~Wu, G.~Dai, Z.~Wu, P.~Zhu, W.~Zhang, K.~W. Yeom,
  K.~Deng, End-to-end automatic differentiation of the coronavirus disease 2019
  (covid-19) from viral pneumonia based on chest ct, European journal of
  nuclear medicine and molecular imaging 47~(11) (2020) 2516--2524.
\newblock \href {http://dx.doi.org/10.1007/s00259-020-04929-1}
  {\path{doi:10.1007/s00259-020-04929-1}}.

\bibitem{a5}
X.~Yu, S.~Lu, L.~Guo, S.-H. Wang, Y.-D. Zhang, Resgnet-c: A graph convolutional
  neural network for detection of covid-19, Neurocomputing\href
  {http://dx.doi.org/10.1016/j.neucom.2020.07.144}
  {\path{doi:10.1016/j.neucom.2020.07.144}}.

\bibitem{a6}
M.~Turkoglu, Covid-19 detection system using chest ct images and multiple
  kernels-extreme learning machine based on deep neural network, IRBM\href
  {http://dx.doi.org/10.1016/j.irbm.2021.01.004}
  {\path{doi:10.1016/j.irbm.2021.01.004}}.

\bibitem{a7}
K.~Gao, J.~Su, Z.~Jiang, L.-L. Zeng, Z.~Feng, H.~Shen, P.~Rong, X.~Xu, J.~Qin,
  Y.~Yang, et~al., Dual-branch combination network (dcn): Towards accurate
  diagnosis and lesion segmentation of covid-19 using ct images, Medical image
  analysis 67 (2021) 101836.
\newblock \href {http://dx.doi.org/10.1016/j.media.2020.101836}
  {\path{doi:10.1016/j.media.2020.101836}}.

\bibitem{a9}
S.-H. Wang, V.~V. Govindaraj, J.~M. G{\'o}rriz, X.~Zhang, Y.-D. Zhang, Covid-19
  classification by fgcnet with deep feature fusion from graph convolutional
  network and convolutional neural network, Information Fusion 67 (2021)
  208--229.
\newblock \href {http://dx.doi.org/10.1016/j.inffus.2020.10.004}
  {\path{doi:10.1016/j.inffus.2020.10.004}}.

\bibitem{a10}
S.~Wang, B.~Kang, J.~Ma, X.~Zeng, M.~Xiao, J.~Guo, M.~Cai, J.~Yang, Y.~Li,
  X.~Meng, et~al., A deep learning algorithm using ct images to screen for
  corona virus disease (covid-19), European Radiology (2021) 1--9\href
  {http://dx.doi.org/10.1007/s00330-021-07715-1}
  {\path{doi:10.1007/s00330-021-07715-1}}.

\bibitem{a11}
X.~Ouyang, J.~Huo, L.~Xia, F.~Shan, J.~Liu, Z.~Mo, F.~Yan, Z.~Ding, Q.~Yang,
  B.~Song, et~al., Dual-sampling attention network for diagnosis of covid-19
  from community acquired pneumonia, IEEE Transactions on Medical Imaging
  39~(8) (2020) 2595--2605.
\newblock \href {http://dx.doi.org/10.1109/TMI.2020.2995508}
  {\path{doi:10.1109/TMI.2020.2995508}}.

\bibitem{a12}
H.~Gunraj, L.~Wang, A.~Wong, Covidnet-ct: A tailored deep convolutional neural
  network design for detection of covid-19 cases from chest ct images,
  Frontiers in medicine 7.
\newblock \href {http://dx.doi.org/10.3389/fmed.2020.608525}
  {\path{doi:10.3389/fmed.2020.608525}}.

\bibitem{a13}
S.~Ahuja, B.~K. Panigrahi, N.~Dey, V.~Rajinikanth, T.~K. Gandhi, Deep transfer
  learning-based automated detection of covid-19 from lung ct scan slices,
  Applied Intelligence 51~(1) (2021) 571--585.
\newblock \href {http://dx.doi.org/10.1007/s10489-020-01826-w}
  {\path{doi:10.1007/s10489-020-01826-w}}.

\bibitem{a14}
M.~Loey, G.~Manogaran, N.~E.~M. Khalifa, A deep transfer learning model with
  classical data augmentation and cgan to detect covid-19 from chest ct
  radiography digital images, Neural Computing and Applications (2020)
  1--13\href {http://dx.doi.org/10.1007/s00521-020-05437-x}
  {\path{doi:10.1007/s00521-020-05437-x}}.

\bibitem{a15}
J.~Pu, J.~Leader, A.~Bandos, J.~Shi, P.~Du, J.~Yu, B.~Yang, S.~Ke, Y.~Guo,
  J.~B. Field, et~al., Any unique image biomarkers associated with covid-19?,
  European radiology 30 (2020) 6221--6227.
\newblock \href {http://dx.doi.org/10.1007/s00330-020-06956-w}
  {\path{doi:10.1007/s00330-020-06956-w}}.

\bibitem{a16}
T.~Goel, R.~Murugan, S.~Mirjalili, D.~K. Chakrabartty, Automatic screening of
  covid-19 using an optimized generative adversarial network, Cognitive
  Computation (2021) 1--16\href {http://dx.doi.org/10.1007/s12559-020-09785-7}
  {\path{doi:10.1007/s12559-020-09785-7}}.

\bibitem{a18}
Z.~Zhu, Z.~Xingming, G.~Tao, T.~Dan, J.~Li, X.~Chen, Y.~Li, Z.~Zhou, X.~Zhang,
  J.~Zhou, et~al., Classification of covid-19 by compressed chest ct image
  through deep learning on a large patients cohort, Interdisciplinary Sciences:
  Computational Life Sciences 13~(1) (2021) 73--82.
\newblock \href {http://dx.doi.org/10.1007/s12539-020-00408-1}
  {\path{doi:10.1007/s12539-020-00408-1}}.

\bibitem{a19}
A.~A. Ardakani, A.~R. Kanafi, U.~R. Acharya, N.~Khadem, A.~Mohammadi,
  Application of deep learning technique to manage covid-19 in routine clinical
  practice using ct images: Results of 10 convolutional neural networks,
  Computers in Biology and Medicine 121 (2020) 103795.
\newblock \href {http://dx.doi.org/10.1016/j.compbiomed.2020.103795}
  {\path{doi:10.1016/j.compbiomed.2020.103795}}.

\bibitem{a20}
Y.~Pathak, P.~K. Shukla, A.~Tiwari, S.~Stalin, S.~Singh, Deep transfer learning
  based classification model for covid-19 disease, Irbm\href
  {http://dx.doi.org/10.1016/j.irbm.2020.05.003}
  {\path{doi:10.1016/j.irbm.2020.05.003}}.

\bibitem{a21}
S.~Hu, Y.~Gao, Z.~Niu, Y.~Jiang, L.~Li, X.~Xiao, M.~Wang, E.~F. Fang,
  W.~Menpes-Smith, J.~Xia, et~al., Weakly supervised deep learning for covid-19
  infection detection and classification from ct images, IEEE Access 8 (2020)
  118869--118883.
\newblock \href {http://dx.doi.org/10.1109/ACCESS.2020.3005510}
  {\path{doi:10.1109/ACCESS.2020.3005510}}.

\bibitem{a22}
A.~Amyar, R.~Modzelewski, H.~Li, S.~Ruan, Multi-task deep learning based ct
  imaging analysis for covid-19 pneumonia: Classification and segmentation,
  Computers in Biology and Medicine 126 (2020) 104037.
\newblock \href {http://dx.doi.org/10.1016/j.compbiomed.2020.104037}
  {\path{doi:10.1016/j.compbiomed.2020.104037}}.

\bibitem{a23}
N.~E.~M. Khalifa, M.~H.~N. Taha, A.~E. Hassanien, S.~H.~N. Taha, The detection
  of covid-19 in ct medical images: A deep learning approach, in: Big Data
  Analytics and Artificial Intelligence Against COVID-19: Innovation Vision and
  Approach, Springer, 2020, pp. 73--90.
\newblock \href {http://dx.doi.org/10.1007/978-3-030-55258-9\_5}
  {\path{doi:10.1007/978-3-030-55258-9\_5}}.

\bibitem{a24}
E.~Matsuyama, et~al., A deep learning interpretable model for novel coronavirus
  disease (covid-19) screening with chest ct images, Journal of Biomedical
  Science and Engineering 13~(07) (2020) 140.
\newblock \href {http://dx.doi.org/10.4236/jbise.2020.137014}
  {\path{doi:10.4236/jbise.2020.137014}}.

\bibitem{a25}
U.~{\"O}zkaya, {\c{S}}.~{\"O}zt{\"u}rk, S.~Budak, F.~Melgani, K.~Polat,
  Classification of covid-19 in chest ct images using convolutional support
  vector machines, arXiv preprint arXiv:2011.05746.

\bibitem{a26}
X.~Deng, H.~Shao, L.~Shi, X.~Wang, T.~Xie, A classification--detection approach
  of covid-19 based on chest x-ray and ct by using keras pre-trained deep
  learning models, Computer Modeling in Engineering \& Sciences 125~(2) (2020)
  579--596.
\newblock \href {http://dx.doi.org/10.32604/cmes.2020.011920}
  {\path{doi:10.32604/cmes.2020.011920}}.

\bibitem{a27}
V.~Bhargavi, R.~D. Rubi, R.~Subramanian, S.~Yadav, Automatic identification of
  covid-19 regions on ct-images using deep learning, European Journal of
  Molecular \& Clinical Medicine 7~(3) (2021) 668--676.

\bibitem{a28}
Z.~A. Khalaf, S.~S. Hammadi, A.~K. Mousa, H.~M. Ali, H.~R. Alnajar, R.~H.
  Mohsin, Coronavirus disease (covid-19) detection using deep features
  learning.

\bibitem{a29}
H.~Swapnarekha, H.~S. Behera, J.~Nayak, B.~Naik, Covid ct-net: A deep learning
  framework for covid-19 prognosis using ct images, Journal of
  Interdisciplinary Mathematics (2021) 1--26\href
  {http://dx.doi.org/10.1080/09720502.2020.1857905}
  {\path{doi:10.1080/09720502.2020.1857905}}.

\bibitem{a30}
E.~D. Carvalho, E.~D. Carvalho, A.~O. de~Carvalho~Filho, F.~H.~D.
  de~Ara{\'u}jo, R.~d. A.~L. Rab{\^e}lo, Diagnosis of covid-19 in ct image
  using cnn and xgboost, in: 2020 IEEE Symposium on Computers and
  Communications (ISCC), IEEE, 2020, pp. 1--6.
\newblock \href {http://dx.doi.org/10.1109/ISCC50000.2020.9219726}
  {\path{doi:10.1109/ISCC50000.2020.9219726}}.

\bibitem{r1}
A.~Sharif~Razavian, H.~Azizpour, J.~Sullivan, S.~Carlsson, Cnn features
  off-the-shelf: an astounding baseline for recognition, in: Proceedings of the
  IEEE conference on computer vision and pattern recognition workshops, 2014,
  pp. 806--813.

\bibitem{r3}
M.~Tan, Q.~Le, Efficientnet: Rethinking model scaling for convolutional neural
  networks, in: International Conference on Machine Learning, PMLR, 2019, pp.
  6105--6114.

\bibitem{no}
M.~Khodatars, A.~Shoeibi, N.~Ghassemi, M.~Jafari, A.~Khadem, D.~Sadeghi,
  P.~Moridian, S.~Hussain, R.~Alizadehsani, A.~Zare, et~al., Deep learning for
  neuroimaging-based diagnosis and rehabilitation of autism spectrum disorder:
  A review, arXiv preprint arXiv:2007.01285.

\bibitem{ns}
D.~Sadeghi, A.~Shoeibi, N.~Ghassemi, P.~Moridian, A.~Khadem, R.~Alizadehsani,
  M.~Teshnehlab, J.~M. Gorriz, S.~Nahavandi, An overview on artificial
  intelligence techniques for diagnosis of schizophrenia based on magnetic
  resonance imaging modalities: Methods, challenges, and future works, arXiv
  preprint arXiv:2103.03081.

\bibitem{ngan}
I.~J. Goodfellow, J.~Pouget-Abadie, M.~Mirza, B.~Xu, D.~Warde-Farley, S.~Ozair,
  A.~Courville, Y.~Bengio, Generative adversarial networks, arXiv preprint
  arXiv:1406.2661.

\bibitem{nngan}
N.~Ghassemi, A.~Shoeibi, M.~Rouhani, Deep neural network with generative
  adversarial networks pre-training for brain tumor classification based on mr
  images, Biomedical Signal Processing and Control 57 (2020) 101678.
\newblock \href {http://dx.doi.org/10.1016/j.bspc.2019.101678}
  {\path{doi:10.1016/j.bspc.2019.101678}}.

\bibitem{nnfive}
M.~Mirza, S.~Osindero, Conditional generative adversarial nets, arXiv preprint
  arXiv:1411.1784.

\bibitem{r9}
P.~Y. Simard, D.~Steinkraus, J.~C. Platt, et~al., Best practices for
  convolutional neural networks applied to visual document analysis., in:
  Icdar, Vol.~3, Citeseer, 2003.

\bibitem{r7}
J.~Howard, S.~Gugger, Deep Learning for Coders with fastai and PyTorch,
  O'Reilly Media, 2020.

\bibitem{nnfour}
T.~Abraham, et~al., Unpaired image-to-image translation,
  \url{https://github.com/tmabraham/UPIT} (2021).

\bibitem{nnthree}
R.~Wightmann, Pytorch image models,
  \url{https://github.com/rwightman/pytorch-image-models/} (2021).

\bibitem{n1}
A.~Shoeibi, N.~Ghassemi, R.~Alizadehsani, M.~Rouhani, H.~Hosseini-Nejad,
  A.~Khosravi, M.~Panahiazar, S.~Nahavandi, A comprehensive comparison of
  handcrafted features and convolutional autoencoders for epileptic seizures
  detection in eeg signals, Expert Systems with Applications 163 (2021) 113788.
\newblock \href {http://dx.doi.org/10.1016/j.eswa.2020.113788}
  {\path{doi:10.1016/j.eswa.2020.113788}}.

\end{thebibliography}

\end{document}